\documentclass[utf8]{FrontiersinHarvard}
\usepackage{url,hyperref,lineno,microtype,booktabs}
\usepackage[singlespacing]{setspace}
\usepackage{accsupp}
\hypersetup{hidelinks}
\def\keyFont{\fontsize{8}{11}\helveticabold}
\def\firstAuthorLast{Hossain {et~al.}}
\def\Authors{Tanzir Hossain\,$^{1}$, Rajib Rana\,$^{2,*}$, Prabal Datta Barua\,$^{2,7}$, Abu Ali Ibn Sina\,$^{3}$, Niall Higgins\,$^{2,4}$, Pascal Elahi\,$^{8}$, Robert Sang\,$^{2}$, and Bj\"orn W. Schuller\,$^{5,6}$}
\def\Address{$^{1}$BRAC University, Dhaka, Bangladesh\\
$^{2}$University of Southern Queensland, Springfield Central, QLD, Australia\\
$^{3}$School of Biotechnology and Biomolecular Sciences, UNSW Sydney, Sydney, NSW, Australia\\
$^{4}$Mental Health and Specialist Services, West Moreton Health, Brisbane, QLD, Australia\\
$^{5}$GLAM--Group on Language, Audio, and Music, Imperial College London, London, United Kingdom\\
$^{6}$Chair of Health Informatics, Technical University of Munich (TUM), Munich, Germany\\
$^{7}$Cogninet AI, Brisbane, QLD, Australia\\
$^{8}$Pawsey Supercomputing Research Centre, Kensington, WA, Australia}
\def\corrAuthor{Rajib Rana}
\def\corrEmail{Rajib.Rana@unisq.edu.au}
\newcommand{\AuthorContributions}{TH contributed to methodology, software development, formal analysis, data curation, visualization, and preparation of the original draft. RR contributed to conceptualization, methodology, project administration, supervision, funding acquisition, interpretation of results, and writing and editing of the manuscript. PDB contributed to methodology, validation, interpretation of results, and review and editing of the manuscript. AAIS contributed to interpretation of the drug-response model and biological context, and review and editing of the manuscript. NH contributed to interpretation of the study, critical review, and editing of the manuscript. PE contributed to high-performance and quantum computing methodology, computational resource support, technical validation, and review and editing of the manuscript. RS contributed to supervision, research direction, interpretation of the findings, and review and editing of the manuscript. BWS contributed to conceptualization, methodological guidance, supervision, interpretation of the quantum-computing results, and review and editing of the manuscript.}
\newcommand{\FundingStatement}{This work was supported by computational resources provided by the Pawsey Supercomputing Research Centre through the Setonix-Q Pilot Scheme under project NCMAS-2026-66, “Closing the Causality Gap: A Computational Framework for Validating Allosteric Drug Mechanisms.” The allocation provided access to Pawsey’s Setonix high-performance computing infrastructure and quantum-enabled computing resources, including access to quantum processing resources through the Quantum Hub Portal and AWS Braket.}
\newcommand{\ConflictStatement}{The authors declare that the research was conducted in the absence of any commercial or financial relationships that could be construed as a potential conflict of interest.}
\newcommand{\DataStatement}{The drug-response data used in this study were obtained from the publicly available AstraZeneca--Sanger Drug Combination DREAM Challenge dataset. The processed optimization instances, quantum hardware output data, noiseless simulation results, classical baseline results, parameter-sensitivity analyses, and scripts used to reproduce the analyses will be made publicly available in a research repository upon publication. Quantum hardware task identifiers and sufficient metadata to reproduce the reported analyses will also be provided, subject to the access conditions of the relevant quantum-computing platforms.}

\newcommand{\TableOneRows}{%
6 & 3 & 2 & 30 & 20 & 176/200 & 88.0 (82.8--91.8) & 5/200  & 100 \\
8 & 4 & 2 & 56 & 56 & 146/200 & 73.0 (66.5--78.7) & 5/200  & 100 \\
9 & 3 & 3 & 72 & 54 & 126/200 & 63.0 (56.1--69.4) & 0/200  & 100 \\
10 & 5 & 2 & 90 & 120 & 141/200 & 70.5 (63.8--76.4) & 0/200  & 100 \\
12 & 4 & 3 & 132 & 162 & 59/200 & 29.5 (23.6--36.2) & 0/200  & 100 \\
15 & 5 & 3 & 210 & 360 & 130/200 & 65.0 (58.2--71.3) & 0/200  & 98 \\
}

\newcommand{\RecoveredRows}{%
6 & 3 & 2 & 07-03 03:33 & 500 & 88.4 (85.3--90.9) & 6\\
6 & 3 & 2 & 07-11 07:10 & 200 & 85.5 (80.0--89.7) & 5\\
8 & 4 & 2 & 07-11 07:12 & 200 & 72.0 (65.4--77.8) & 5\\
10 & 5 & 2 & 07-11 07:19 & 200 & 69.5 (62.8--75.5) & 4\\
12 & 6 & 2 & 07-11 07:25 & 200 & 72.5 (65.9--78.2) & 4\\
14 & 7 & 2 & 07-11 07:38 & 200 & 62.0 (55.1--68.4) & 3\\
15 & 5 & 3 & 07-11 07:48 & 200 & 54.5 (47.6--61.3) & 0\\
18 & 6 & 3 & 07-11 06:06 & 200 & 42.0 (35.4--48.9) & 0\\
18 & 6 & 3 & 07-11 07:55 & 200 & 49.5 (42.6--56.4) & 1\\
18 & 6 & 3 & 07-12 13:37 & 200 & 46.0 (39.2--52.9) & 1\\
20 & 4 & 5 & 07-11 08:09 & 200 & 20.0 (15.0--26.1) & 0\\
20 & 4 & 5 & 07-12 13:47 & 200 & 27.0 (21.3--33.5) & 0\\
21 & 7 & 3 & 07-11 06:09 & 200 & 44.0 (37.3--50.9) & 1\\
21 & 7 & 3 & 07-11 08:02 & 200 & 40.5 (33.9--47.4) & 0\\
21 & 7 & 3 & 07-12 13:40 & 200 & 35.0 (28.7--41.8) & 0\\
25 & 5 & 5 & 07-11 08:17 & 200 & 13.0 (9.0--18.4) & 0\\
25 & 5 & 5 & 07-12 13:54 & 200 & 15.0 (10.7--20.6) & 0\\
30 & 6 & 5 & 07-11 08:23 & 200 & 6.5 (3.8--10.8) & 0\\
35 & 7 & 5 & 07-11 08:33 & 200 & 3.5 (1.7--7.0) & 0\\
}

\newcommand{\Fset}{\mathcal{F}}
\newcommand{\altfigure}[3]{\BeginAccSupp{method=escape,ActualText={#2}}\includegraphics[width=#3]{#1}\EndAccSupp{}}

\begin{document}
\onecolumn
\firstpage{1}
\title[Feasibility and optimum recovery]{Feasibility and optimum recovery in warm-start quantum optimization for a drug-response model on a trapped-ion processor}
\author[\firstAuthorLast]{\Authors}
\address{}
\correspondance{}
\extraAuth{}
{\setlength{\parindent}{0pt}\maketitle}
\newcommand{\MainTextWords}{3279}

\noindent\textbf{Article type:} Brief Research Report\quad
\textbf{Main-text words:} \MainTextWords\quad
\textbf{Figures:} 2\quad\textbf{Tables:} 2

\begin{abstract}
\section{}
On a seven-compound drug-response model, warm-start quantum approximate optimization (QAOA) on IonQ Forte-1 returned valid assignments more often than random bitstrings, but this alone did not show effective optimization. Ideal QAOA raised optimum probability above uniform feasible sampling in only four of twelve reference circuits. Hardware often fell below its own noiseless circuits, while greedy search solved all hardware models within 200 objective evaluations. Expanded simulations showed a gain over feasible sampling in 28 of 35 CAMA-1 panels and none of four 647-V panels. Annealing solved all these panels in every seed. A Grover mixer preserved feasibility and improved optimum probability over feasible sampling in all ten tested models. We analyzed 25 completed tasks containing 5,300 shots from 18 circuits and 14 instances. The encodings use 6--35 qubits and at most 4,900 feasible assignments, which we enumerated to establish exact optima. Noiseless references now cover both the original six circuits and six wider circuits. At 35 qubits, with 4,900 feasible assignments, ideal feasibility was 35.85\%, compared with 7 of 200 valid hardware outputs. Its ideal optimum probability was below both sampling controls. The circuits sample assignments in a model built from measured single-agent and pairwise responses.

\tiny
\keyFont{\section{Keywords:} quantum approximate optimization, warm start, trapped-ion quantum computing, feasibility, classical baselines, drug-response modeling, constrained optimization}
\end{abstract}

\section{Introduction}
Most quantum optimization studies ask whether a quantum circuit produces better samples than an unconstrained random baseline. For constrained problems, however, improved feasibility need not imply improved optimization. We investigate this distinction using archived trapped-ion executions of warm-start QAOA for a dose-assignment model, linking the initial distribution, ideal circuit behavior, hardware execution, exact feasible optima and constructive classical search.

QAOA alternates cost evolution and mixing \citep{farhi2014}; warm-start methods use classical information to set the initial state and sometimes the mixer \citep{egger2021}. Constraint-preserving mixers instead restrict evolution to feasible assignments \citep{hadfield2019}. Prior studies report warm-start improvements for portfolio and MaxCut models \citep{egger2021}, trapped-ion optimization through 40 qubits \citep{pagano2020}, and separate feasibility and quality comparisons with uniform feasible sampling at 14 and 20 qubits \citep{niroula2022}. Drug-combination QUBOs have also been studied with simulated quantum and classical annealing \citep{ramos2026}. These comparisons motivate controls that account for both constraints and objective quality \citep{koch2026,sankar2024}.

We connect the initial product distribution, unchanged noiseless circuits and 25 hardware executions to exact optima and classical search in an AstraZeneca--Sanger DREAM drug-response model \citep{menden2019}. Ideal QAOA beats its initial distribution in eleven of twelve reference circuits but beats uniform feasible sampling in only four. Four additional panels test whether sampling gains extend to different optima. Greedy search solves every hardware-associated problem at the archived weight; further simulations test all CAMA-1 panels, a second cell line, other score weights and a feasibility-preserving mixer. The largest instance uses 35 qubits for only 4,900 feasible assignments and is fully enumerable. Circuit width describes the encoding and hardware resources, while the feasible-set size describes the search space.

\section{Method}
\subsection{Source data and instance construction}
The optimization model is derived from the AstraZeneca--Sanger Drug Combination DREAM dataset, an experimental screen used to study responses to anticancer compounds and their combinations \citep{menden2019}. Its single-agent dose-response parameters describe how cell viability changes with concentration, while pairwise synergy scores describe interactions between compounds. We use these measurements for the single-compound and pair-interaction terms, respectively.

We use the archived seven-compound panel for the CAMA-1 cell line. The source pipeline retained this fully measured group: synergy observations are available for all 21 compound pairs, avoiding the need to fill missing pair coefficients. It retains observations passing the dataset's quality check, averages repeated pairwise synergy scores, and takes the median single-agent dose-response parameters. The compound identifiers are AKT\_1, BCL2\_BCL2L1, FASN, MAP2K\_1, MTOR\_1, PIK3CB\_PIK3CD and SLC16A4. Subsets of three to seven compounds and grids of two, three or five nonzero dose choices define the hardware-associated instances. This provides experimentally informed optimization coefficients while keeping the feasible sets small enough to enumerate and establish exact reference optima.

\subsection{Dose-assignment representation and score}
For $n$ compounds and $D$ nonzero dose choices, $q=nD$ bits encode $x_{id}\in\{0,1\}$. A compound is absent when its entire block is zero; a block with two or more ones is invalid. A feasible assignment selects two or three compounds, at most one dose per selected compound, and at least two recorded mechanism classes. All compounds within each included instance have distinct class labels, so the diversity check is redundant once the size constraint holds. The archived constraints concern assignment structure; they include no exposure cap or clinical safety rule.

The scalar score is
\begin{equation}
 f(x)=\sum_{i,d}(h_{id}-\lambda t_{id})x_{id}
       +\sum_{i<j}\sum_{d,e}J_{id,je}x_{id}x_{je},\qquad\lambda=0.6.
 \label{eq:score}
\end{equation}
The single-compound term $h_{id}$ is an inhibition estimate from the fitted Hill dose-response curve. Dose choices are fractions of each compound's maximum tested concentration. The original series uses $(0.1,1)$ or $(0.1,0.32,1)$; the larger instances also use archived five-level grids. The pairwise term multiplies the measured synergy coefficient by the product of the dose fractions. This assumes a dose dependence that was not measured as a response surface. The term $t_{id}$ is a class-weighted dose proxy, rather than a toxicity measurement. We retain the archived coefficients throughout. Scores for three-drug combinations extrapolate from single-agent and pairwise terms and have no calibrated clinical interpretation.

For the original six circuits, binary-state and compound-assignment enumeration independently identify the feasible set $\Fset$. For larger instances, we enumerate two- and three-compound assignments directly and separately decode every observed bitstring, avoiding allocation of all $2^q$ states. Because each mechanism class is distinct,
\begin{equation}
 |\Fset|=\binom{n}{2}D^2+\binom{n}{3}D^3.
 \label{eq:count}
\end{equation}
The original six instances contain 20--360 feasible assignments; across the expanded set the maximum is 4,900 at 35 qubits. We find each optimum by evaluating Equation~\ref{eq:score} over the full feasible set.

\subsection{Quantum hardware and execution}
We analyze circuits executed on the IonQ Forte-1 trapped-ion quantum processor through Amazon Braket. Forte encodes qubits in trapped ions, uses laser-driven operations, and supports interactions between any pair of qubits \citep{chen2024}. This connectivity accommodates the pairwise terms of the cost Hamiltonian. Our experiments use 6--35 qubits for 20--4,900 feasible assignments, with one qubit per compound-dose choice. The submitted programs contain single-qubit rotations and CNOT gates; their CNOT counts range from 30 to 1,190. These counts describe the submitted circuits, since native compiled gate counts are unavailable.

Each shot executes a circuit and measures its qubits to produce one candidate binary string, which we decode into compound and dose choices. We analyze 25 hardware executions of 18 distinct circuits representing 14 optimization instances: 24 executions contain 200 shots and one contains 500, for 5,300 measurements in total. We analyze identical-circuit repetitions separately. Six circuits at 6--15 qubits provide the original reference series (Table~\ref{tab:hardware}); the additional instances reach 35 qubits and 4,900 feasible assignments (Table~\ref{tab:expanded}). We compare each execution with classical controls and noiseless calculations of the same submitted circuit. The supplement gives circuit details and task identifiers.

\subsection{Executed circuit and noiseless reconstruction}
The recorded initial state is a product state,
\begin{equation}
 |\psi_0\rangle=\bigotimes_{v=0}^{q-1}R_y(\theta_v)|0\rangle,
 \qquad P_0(x)=\prod_v c_v^{x_v}(1-c_v)^{1-x_v},\quad
 c_v=\sin^2(\theta_v/2).
 \label{eq:prior}
\end{equation}
The archived one-layer circuit applies a diagonal Ising cost unitary followed by local warm-start mixers:
\begin{equation}
 |\psi\rangle=\left[\bigotimes_v R_y(\theta_v)R_z(2\beta)R_y(-\theta_v)\right]
 \exp\!\left[-i\gamma\left(\sum_v a_vZ_v+\sum_{v<w}b_{vw}Z_vZ_w\right)\right]|\psi_0\rangle.
 \label{eq:circuit}
\end{equation}
Products act from right to left. We recover the cost coefficients and angles from the archived runners and check them against the submitted programs. The cost includes penalties for invalid assignments. Mixing consists of single-qubit rotations; the CNOT--$R_z$--CNOT blocks implement diagonal cost interactions. These circuits contain no XY exchange mixer and do not preserve feasibility by construction.

We retain the archived hardware parameters. For the original six circuits, we check a NumPy gate simulator and an independent calculation of Equation~\ref{eq:circuit} against PennyLane. We first extend the latter calculation to four circuits from 18 through 25 qubits, forming phases in chunks and updating amplitude pairs in place. Each width has one distinct circuit shared by its repeated tasks.

We then calculate full statevectors at 30 qubits (2,875 feasible assignments) and 35 qubits (4,900 feasible assignments) on a Pawsey node with sufficient memory. We use single-precision complex amplitudes, initialize them in Gray-code order and apply local mixers in parallel. Against double precision at 25 qubits, the feasible probability differs by less than $10^{-7}$. The accelerated implementation also reproduces all six original distributions. We retain the recorded qubit order, with $q[0]$ as the first outcome bit. The supplement gives precision checks, memory requirements and transferred-parameter details.

\subsection{Classical controls and performance measures}
We use uniform sampling over all $2^q$ bitstrings as an encoding-space reference. Its feasibility is calculated exactly as $|\Fset|/2^q$. The warm-start-only control samples the independent Bernoulli distribution in Equation~\ref{eq:prior}; it is classically implementable and isolates the distribution already present before cost evolution and mixing. Uniform feasible sampling draws directly from $\Fset$ without evaluating the objective: choose a size $m\in\{2,3\}$ with probability proportional to $\binom{n}{m}D^m$, choose a subset uniformly, and choose its doses uniformly. Every draw is feasible. We use exhaustive enumeration as an exact solver, with one objective evaluation per feasible assignment.

For each distribution we calculate feasible probability $P_{\Fset}$ and probability $P_\star$ of any globally optimal feasible assignment. For the analytically known distributions, the probability of at least one optimum in $B$ independent draws is
\begin{equation}
 P_{\mathrm{hit}}(B)=1-(1-P_\star)^B,\qquad B=200.
 \label{eq:budget}
\end{equation}
We use this expression for the analytic sampling controls.

Multiplying each recorded outcome probability by the task's shot count gives an integer count within $10^{-7}$ in every case, and counts sum to the task's declared 200 or 500 shots. Feasibility uses all shots as the denominator, including invalid outputs. We report pointwise 95\% Wilson intervals for hardware proportions \citep{wilson1927}.

To distinguish near-optimal from poor feasible outputs, we normalize scores as $u(x)=[f(x)-f_{\min}]/[f_\star-f_{\min}]$, where the minimum and maximum are over $\Fset$. We report the mean $u$ among valid hardware shots and the observed best-sample gap $1-\max u$; invalid outputs remain represented by the separate feasibility metric. Both summaries describe the observed batch.

We also implement deterministic greedy addition followed by feasible local improvement. Starting from the empty assignment, the search adds the best available compound-dose choice until two compounds are selected. It then considers dose changes, additions, deletions, and compound swaps, moving to the best improving neighbor. A cache avoids repeated objective evaluations, with a cap of 200 unique evaluations including partial construction states. The search stops at the cap or when no neighbor improves the score. We compare the search result with the exact optimum only after search.

We also run simulated annealing from a random feasible assignment, using the same neighborhood, geometric cooling and a cap of 200 unique evaluations for each of 100 seeds. We report the fraction reaching the exact optimum and median evaluations to first hit among successful seeds; the proposal limit and temperature schedule are specified in the supplement.

We retain $\lambda=0.6$ because it is the archived score weight and sweep 0.3--0.9 in steps of 0.15, re-enumerating optima, rescoring fixed distributions and rerunning both searches. The hardware models retain their optima and control rankings across this sweep, while retrained panel rankings change (Supplementary Table~S9).

For the six original instances and four original panels, we simulate one-layer Grover-mixer QAOA from the uniform feasible state, using the unpenalized score and a selective phase shift on that state \citep{bartschi2020}. We train both angles with the same three seeds and Adam settings used for the panel simulations; feasibility is one by construction.

\subsection{Panel selection and initialization}
For the additional simulations, we draw four distinct four-compound panels uniformly without replacement from the 35 possible panels in the seven-compound CAMA-1 model, using random seed 20260905. We fix panel membership before optimization. Each panel uses three dose levels, 12 qubits, and the same score weight and constraints. Panel selection does not depend on the simulated results.

We construct the penalty Hamiltonian and product prior for each panel, then train one-layer QAOA with initialization seeds 0, 1 and 2. Each run uses 60 Adam updates and a learning rate of 0.1. We retain the final iterate from every run. A NumPy implementation computes exact statevector gradients, which we check against PennyLane. The original 12 simulations ran on a Pawsey Setonix ARM node. We extend the study to all 35 CAMA-1 panels and four panels from 647-V, selected by measurement coverage before optimization. We also retrain the original four panels at each score weight. These added runs use the same implementation locally; we retain every final iterate. The supplement gives panel membership, the deterministic second-cell-line selection rule and all results.

\section{Results}
\subsection{Original six-task reference: feasibility}
Hardware returns valid assignments more often than uniform random bitstrings in every reference instance. Hardware feasibility ranges from 29.5\% to 88.0\%, compared with 1.10--31.25\% for random bitstrings. The submitted circuits contain 30--210 CNOT gates (Table~\ref{tab:hardware}; Supplementary Table~S1).

The strongest enrichment mainly reflects the sparse encoding. At 15 qubits, only 1.10\% of bitstrings are feasible, so a $59.2\times$ enrichment still leaves over a third of hardware shots invalid. A uniform feasible sampler returns a valid assignment on every draw (Figure~\ref{fig:feasibility}A).

Most reference hardware batches also improve feasibility over initialization, but the 12-qubit exception is already present in the ideal circuit. Hardware exceeds the warm-start-only control in five cases. At 12 qubits, ideal feasibility is 31.18\%, below the prior's 32.51\%. Across the reference circuits, ideal feasibility ranges from 31.18\% to 89.96\%, while the prior ranges from 31.68\% to 62.42\% (Supplementary Table~S1).

\subsection{Optimum recovery in the reference subset}
Hardware recovers the optimum in only two of the six reference batches. We independently find the same unique optimum in all six models: the same three compounds at their highest encoded dose, with score $1.769364$. Hardware returns it five times each at 6 and 8 qubits and never in the other four batches. A zero count gives a Wilson upper endpoint of approximately 1.88\% for the per-shot probability; it does not establish zero probability (Table~\ref{tab:hardware}).

\begin{figure}[t]
\centering
\altfigure{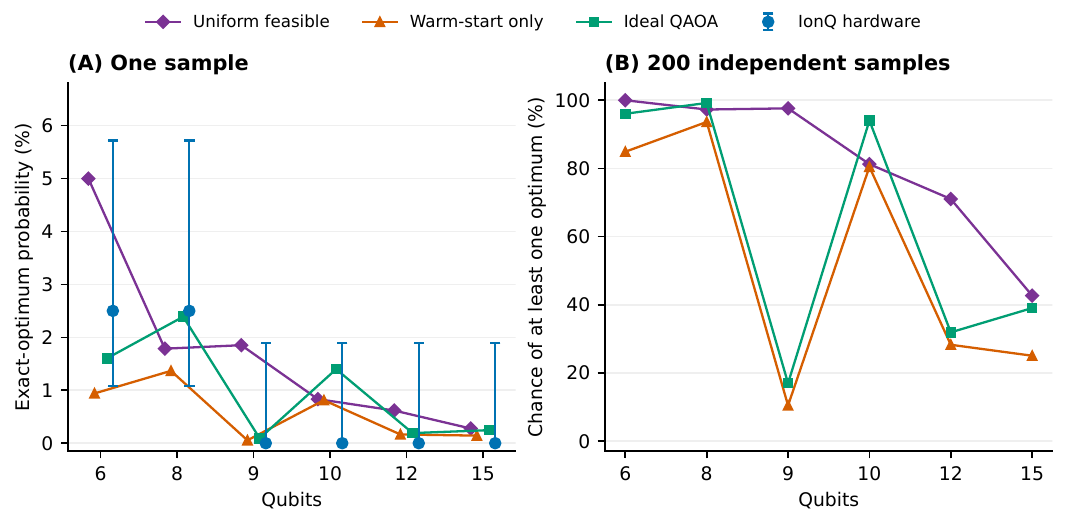}{Panel A shows exact optimum probabilities and hardware counts with uncertainty. Hardware observes an optimum only at six and eight qubits. Panel B shows calculated 200-draw optimum recovery for the analytic controls. At 15 qubits the rates are 42.7 percent for uniform feasible sampling, 39.0 percent for ideal QAOA, and 25.0 percent for the initial product distribution.}{\textwidth}
\caption{Optimum recovery with appropriate controls. \textbf{(A)} Per-shot optimum probability; hardware error bars are pointwise 95\% Wilson intervals, including an upper bound when no optimum is observed. Markers are offset slightly for visibility. \textbf{(B)} Calculated probability of at least one optimum in 200 independent samples from the three analytic distributions, using Equation~\ref{eq:budget}. Panel B uses analytic probabilities and assumes independent draws.}
\label{fig:optimum}
\end{figure}

High feasibility can accompany poor optimum recovery. At 15 qubits, hardware returns 130 valid outputs but no optimum. Ideal optimum probability is 0.247\%, compared with 0.144\% for the prior and 0.278\% for uniform feasible sampling. Their calculated 200-draw success probabilities are 39.0\%, 25.0\% and 42.7\%, respectively; uniform bitstrings give only 0.61\% (Figure~\ref{fig:optimum}).

Ideal QAOA improves on initialization in every reference circuit, but it beats uniform feasible sampling only at 8 and 10 qubits. The 10-qubit ideal circuit predicts a 94.0\% chance of an optimum within 200 draws, yet hardware returns none. The later execution uses different parameters and cannot resolve this discrepancy (Figure~\ref{fig:optimum}; Supplementary Table~S2).

\subsection{Solution quality and a constructive search baseline}
Greedy local search solves all six reference instances within 21--81 unique objective evaluations. Hardware's best-sample gaps instead range from zero to 39.5\% of the feasible score range. The four batches without an optimum have gaps of 18.8--39.5\% (Supplementary Table~S4 and Figure~S1).

Across all 14 hardware models at $\lambda=0.6$, annealing reaches the optimum in 15--100\% of seeds within the evaluation cap. On the original six, success is 98--100\%, with median first-hit counts of 11--87.5 evaluations among successful seeds. Greedy solves all 14 in 21--200 evaluations, whereas exact enumeration uses 20--4,900. At 35 qubits, with 4,900 feasible assignments, annealing succeeds in 15 of 100 seeds and greedy reaches the optimum at its cap (Table~\ref{tab:hardware}; Supplementary Table~S8).

\subsection{Expanded hardware records and repeated executions}
Optimum recovery remains uneven across the additional instances. Hardware finds an optimum in the 10--14-qubit cases, in two of three 18-qubit executions and in one of three 21-qubit executions. It finds none in the five-level dose instances, including the 35-qubit encoding with 4,900 feasible assignments (Table~\ref{tab:expanded}).

Repeated executions show appreciable variation for some fixed circuits. Feasibility ranges from 35.0\% to 44.0\% at 21 qubits and from 42.0\% to 49.5\% at 18 qubits. The two 6-qubit runs are closer, at 88.0--88.4\%, with 200 and 500 shots. We retain the additional circuits with changed parameters or compound membership as separate records (Supplementary Figure~S2).

\begin{figure}[h!]
\centering
\altfigure{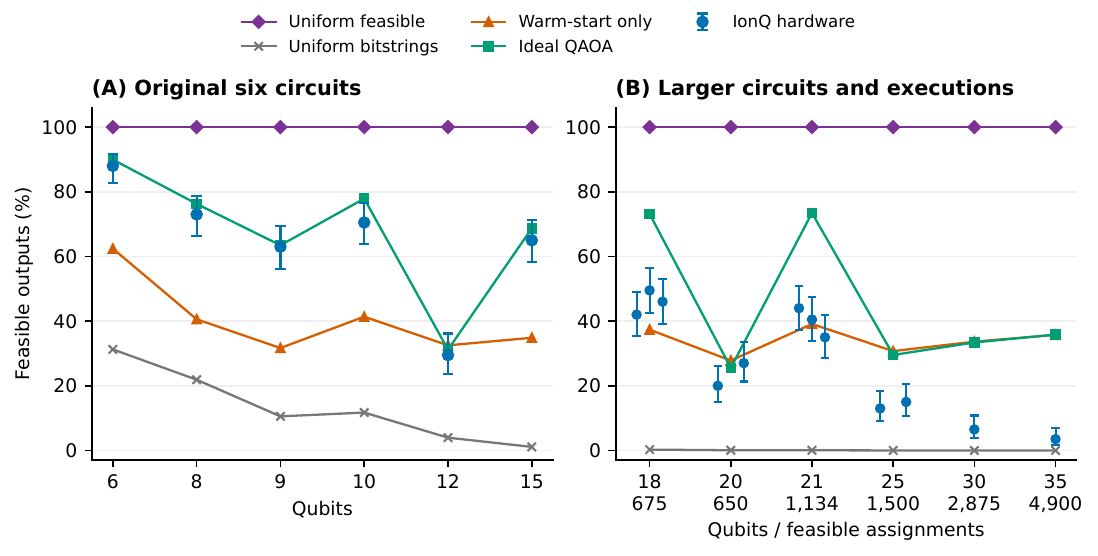}{Panel A compares feasibility for the original six circuits. Panel B compares individual hardware executions, initial product distributions and ideal circuits through 35 qubits with 4,900 feasible assignments. The last two ideal values are 33.39 and 35.85 percent, compared with hardware values of 6.5 and 3.5 percent. Uniform feasible sampling is always valid.}{\textwidth}
\caption{Feasibility with classical and noiseless controls. \textbf{(A)} Original six circuits. \textbf{(B)} Larger circuits, with individual hardware executions offset horizontally and feasible-space sizes below qubit counts. Hardware error bars are pointwise 95\% Wilson intervals. Other series are numerically evaluated probabilities; repeated tasks share one ideal circuit reference. Lines guide the eye across different instances and do not imply a scaling law.}
\label{fig:feasibility}
\end{figure}

Hardware loses feasible probability relative to its own ideal circuit at 18, 21 and 25 qubits. Every pointwise hardware interval at these settings excludes the ideal value, while both 20-qubit intervals include it. At 18 and 21 qubits, ideal feasibility is 73.16\% and 73.44\%, respectively. The 20- and 25-qubit ideal circuits already reduce feasibility below their priors; hardware adds a further shortfall at 25 qubits (Figure~\ref{fig:feasibility}B; Supplementary Tables~S6--S7).

Ideal optimum recovery improves over initialization at these four settings but exceeds uniform feasible sampling only at 18 and 21 qubits. Ideal optimum probability ranges from 0.0239\% to 0.1666\%. Mean normalized quality among valid outputs is 39.0--51.0\% ideally and 37.7--50.0\% on hardware. These conditional means exclude invalid outputs (Supplementary Tables~S6--S7).

The two widest circuits show an execution shortfall alongside weak ideal optimization. At 30 qubits, with 2,875 feasible assignments, ideal feasibility is 33.39\%, close to the prior's 33.64\%, while hardware returns only 13 of 200 valid outputs. At 35 qubits, with 4,900 feasible assignments, ideal feasibility is 35.85\%, close to the prior's 35.75\%, while hardware returns 7 of 200. The latter batch has a Wilson interval of 1.7--7.0\% and a best normalized gap of 40.4\%. Both ideal optimum probabilities fall below uniform feasible sampling; at 35 qubits, with 4,900 feasible assignments, the ideal probability of 0.01197\% also falls below the prior's 0.01217\% (Figure~\ref{fig:feasibility}B; Supplementary Tables~S6--S7).

\subsection{Results for the additional panels}
Ideal QAOA beats both sampling controls in every run of the four-panel simulation. Each panel has 162 feasible assignments and a different unique optimum. Across 12 panel/seed combinations, ideal optimum probability is 2.23--3.43\%, compared with 1.26--1.48\% for initialization and 0.617\% for uniform feasible sampling. Greedy search still solves each panel in 36--50 evaluations (Supplementary Table~S3).

The wider panel study shows that the sampling gain depends on the model. Ideal QAOA beats uniform feasible sampling in all three seeds for 28 of 35 CAMA-1 panels (80\%) and for none of four 647-V panels. Greedy solves all CAMA-1 panels and three of four 647-V panels (75\%); annealing solves every panel in all 100 seeds (Supplementary Table~S11 and Figure~S3).

Weight sensitivity leaves the archived sampling comparison unchanged but affects retraining and search. All 14 hardware models retain the same optimum across the sweep, while greedy recovers it in 61 of 70 model/weight combinations. Across the original four panels and five weights, retrained QAOA beats feasible sampling in 52 of 60 runs and its product prior in 59 of 60 (Supplementary Table~S9).

\subsection{Feasibility-preserving mixer results}
The Grover mixer removes invalid outcomes and beats uniform feasible sampling in every tested run. Across the six original instances and four original panels, optimum probability ranges from 1.17\% to 27.24\%. It exceeds the archived penalty-QAOA probability on all six reference instances but falls below the trained penalty-QAOA probability on all four panels. Greedy solves all ten models (Supplementary Table~S10).

\section{Discussion}
Warm-start QAOA improves some sampling probabilities in this model, but feasibility enrichment alone does not establish effective optimization. Ideal QAOA beats initialization in eleven of twelve reference circuits and uniform feasible sampling in only four. The full panel study narrows that gain to 28 of 35 CAMA-1 panels and none of four 647-V panels. At the archived weight, greedy solves all hardware models and CAMA-1 panels, while annealing also solves the 647-V panel that greedy misses. These results support no computational-advantage claim.

The noiseless references separate circuit limitations from execution shortfalls. The 20- and 25-qubit ideal circuits slightly reduce feasibility relative to initialization. Hardware loses much of the ideal feasible mass at 18 and 21 qubits and compounds the circuit limitation at 25 qubits. The new references at 30 qubits (2,875 feasible assignments) and 35 qubits (4,900 feasible assignments) show the same combination: little ideal optimization gain and a substantial loss of feasible probability during execution. Preparation controls, calibration records and native compiled circuits would be needed to identify the source of an execution discrepancy.

The resource comparisons have limited scope. Compounds, dose grids and variational parameters change alongside width, so we cannot attribute performance differences to qubit count. Pointwise Wilson intervals assume independent shots within one execution and do not capture drift between runs. We do not extrapolate future hardware batch success from a single histogram. Objective-evaluation counts exclude quantum training and execution costs and cannot establish an end-to-end timing comparison.

Our comparisons extend prior work separating feasibility from conditional quality in trapped-ion optimization \citep{niroula2022} and comparing drug-combination models with classical annealing \citep{ramos2026}. The product prior uses model scores, so relaxation-based warm-start guarantees do not automatically apply \citep{egger2021}. The Grover-mixer simulations confirm that feasibility can be preserved \citep{hadfield2019,bartschi2020}, but improved optimum probability over penalty QAOA depends on the panel. Because initialization and training also change, this comparison does not isolate the mixer alone. We construct the feasible state directly in simulation; its hardware preparation cost remains to be assessed. Objective-aware classical controls remain essential \citep{koch2026}.

The biological interpretation remains limited by the model. The 647-V panels add a second cell-line model and show that the CAMA-1 sampling pattern does not transfer automatically. Overlapping panels within each cell line are not independent biological replications, and the original six instances share one optimum. Dose-scaled pairwise synergy extrapolates beyond measured response surfaces, while the class-weighted dose term is a proxy rather than measured toxicity. These coefficients define a computational score, not a validated three-drug response or a treatment-safety assessment.

\section*{Data availability statement}
\DataStatement

\section*{Author contributions}
\AuthorContributions

\section*{Funding}
\FundingStatement

\section*{Conflict of interest}
\ConflictStatement

\section*{Acknowledgments}
The authors have no additional acknowledgments.

\bibliographystyle{Frontiers-Harvard}
\bibliography{references}

\clearpage
\begin{table}[t]
\caption{Hardware observations for the original six-task series. Each instance has 200 shots. $q$: qubits; $n$: compounds; $D$: nonzero dose choices; CX: submitted CNOT gates; $|\Fset|$: exactly enumerated feasible assignments. Parentheses contain pointwise 95\% Wilson intervals. The optimum column counts globally optimal feasible outputs. SA: percentage of 100 annealing seeds reaching the optimum within 200 unique objective evaluations.}
\label{tab:hardware}
\centering
\small
\setlength{\tabcolsep}{3pt}
\begin{tabular}{rrrrrrlrr}
\toprule
$q$ & $n$ & $D$ & CX & $|\Fset|$ & Valid shots & Valid \% (95\% CI) & Optimum & SA (\%)\\
\midrule
\TableOneRows
\bottomrule
\end{tabular}
\end{table}

\begin{table}[h!]
\caption{The 19 additional hardware tasks. Task creation dates/times are UTC in 2026. $q$: qubits; $n$: compounds; $D$: nonzero doses; $N$: shots. Valid proportions have pointwise 95\% Wilson intervals. Opt.: observed globally optimal outputs. Rows with the same instance represent separate completed tasks.}
\label{tab:expanded}
\centering\small
\setlength{\tabcolsep}{5pt}
\begin{tabular}{rrrlrlr}
\toprule
$q$ & $n$ & $D$ & UTC date/time & $N$ & Valid \% (95\% CI) & Opt.\\
\midrule
\RecoveredRows
\bottomrule
\end{tabular}
\end{table}

\clearpage
\end{document}


\onecolumn
\firstpage{1}
\title[Supplementary Material]{{\helveticaitalic{Supplementary Material}}}
{\setlength{\parindent}{0pt}\maketitle}

\section{Hardware dataset}
The dataset contains 25 completed IonQ Forte-1 tasks. The supplied files include a duplicate of one 10-qubit task, which we count once. Twenty-four tasks have 200 shots and one 6-qubit task has 500 shots, for a total of 5,300. Each record includes measured-qubit order, submitted OpenQASM, output probabilities and task metadata. Every circuit matches an archived instance definition, and all 25 tasks are included.

This is a retrospective analysis of completed hardware executions and was not preregistered. The hardware records are distinct from the source drug-response dataset used to construct the optimization instances. Numerical figures are generated programmatically from the analyzed records.

Main Table~1 and Figures~1A and 2 report the original six-task series, for which we independently checked the noiseless calculations. Figure~1B adds larger-width noiseless and hardware comparisons. Main Table~2 reports the other 19 tasks. In total, the records represent 18 distinct circuits and 14 optimization instances. The dataset also includes the 30- and 35-qubit executions listed in the main paper.

The reproducibility package contains the original JSON files, four archived runner scripts, their instance definitions and SHA-256 hashes. We extract the literal instance arrays without executing the submission code. The complete task UUID inventory appears below. 

\section{Circuit identity and independent calculations}
We extract literal instance arrays from the four archived runners using Python's abstract syntax tree. No hardware-submission code is imported or executed. For each instance, we reconstruct its initial $R_y$ rotations, cost $R_z$ rotations and CNOT pairs, and final $R_y(-\theta)$, $R_z(2\beta)$, $R_y(\theta)$ mixers. We compare every operation, wire, and numeric angle to the submitted OpenQASM, using $10^{-10}$ absolute tolerance for angles. The compiled native program is absent from the task records, so native-gate counts and noise parameters cannot be recovered from these files.

A NumPy simulator applies the submitted gate sequence directly. A second calculation applies the diagonal Ising phase followed by tensor-product mixers. Their full probability vectors agree within $10^{-10}$ maximum absolute error for the original six instances. A PennyLane \texttt{default.qubit} execution of the same gate list provides a third check, agreeing within $10^{-12}$ absolute tolerance. Bell-state and single-wire rotation tests check the bit-order convention. The first character of each outcome string represents wire zero throughout.

Feasibility is independently constructed both by binary-state decoding and by enumerating compound-level assignments. The reconstructed evaluator also matches the original pure decoding, feasibility, and scoring functions on all 38,720 binary states across the original six instances. A state's score is compared only when it is feasible. Each original instance's optimum is recomputed, including ties; all six have exactly one optimum. For the expanded set, feasible assignments are generated directly by compound subsets and dose choices, independently checked against Cartesian assignment enumeration, and scored using both bit-level and assignment-level implementations. This remains small even at 35 qubits (4,900 feasible assignments). Every observed output is also decoded independently. Additional noiseless calculations through 35 qubits (4,900 feasible assignments) are described below. Source coefficients are used at their archived precision.

\section{Classical sampling and uncertainty}
The 10-qubit prior feasibility is 41.36\% when truncated to two decimal places; Table~S1 rounds this value to 41.37\%.

To sample uniformly from the feasible set, choose $m=2$ or $3$ with probability proportional to $\binom{n}{m}D^m$, choose an $m$-compound subset uniformly, and draw one of $D$ doses independently for each chosen compound. Because every included compound has a distinct mechanism class, this procedure returns every feasible assignment with equal probability. This procedure uses the constraints alone.

For warm-start-only sampling, draw each bit independently with probability $\sin^2(\theta_v/2)$. We sum probabilities over the exact feasible and optimal sets instead of using Monte Carlo estimates for these classical distributions. The ideal-QAOA control is similarly analytic.

For $k$ successes in $N$ shots, the reported 95\% Wilson interval is
\begin{equation}
 \frac{\hat p+z^2/(2N)\ \pm\ z\sqrt{\hat p(1-\hat p)/N+z^2/(4N^2)}}{1+z^2/N},
 \quad \hat p=k/N,\quad z=1.9599639845.
\end{equation}
These are pointwise intervals conditional on independent shots within a run. They do not account for unrecorded calibration changes, provider processing, or between-run variability. No multiple-comparison inference is claimed.

The main fixed-budget metric is $1-(1-P_\star)^{200}$ for analytically known distributions. It describes 200 independent draws with replacement. Hardware columns report each actual batch separately. Conditional hardware score summaries are reported in Tables~S4--S5. The code also saves the expected normalized best-of-200 score for the analytic controls as a diagnostic.

\section{Exact control probabilities}
\begin{table}[h!]
\caption{Feasible probability (percent), calculated exactly for analytic controls. Hardware proportions are reconstructed from archived task records. The feasible classical sampler has 100\% feasibility in every instance.}
\centering\small
\begin{tabular}{rrrrrr}
\toprule
Qubits & Uniform bits & Initial product & Ideal QAOA & Hardware & Uniform feasible\\
\midrule
\SupplementFeasRows
\bottomrule
\end{tabular}
\end{table}

\begin{table}[h!]
\caption{Probability of at least one optimum in 200 independent draws (percent), calculated for analytic controls. Observed hardware batches are reported separately.}
\centering\small
\begin{tabular}{rrrrr}
\toprule
Qubits & Uniform bits & Initial product & Ideal QAOA & Uniform feasible\\
\midrule
\SupplementHitRows
\bottomrule
\end{tabular}
\end{table}

\section{Model coefficients and circuit parameters}
The supplied \texttt{CAMA1c7\_D3\_meta.json} and coefficient arrays define the seven-compound CAMA-1 model used for the sensitivity panels. The source builder uses QA-passing AstraZeneca--Sanger DREAM observations, averages pairwise synergy values, and summarizes monotherapy Hill parameters by their median. Its single-agent term is $h_{id}=[100-v_i(c_d)]/100$, with
\begin{equation}
v_i(c)=E_{\infty,i}+\frac{100-E_{\infty,i}}{1+(c/IC_{50,i})^{H_i}}.
\end{equation}
The pair term is $J_{id,je}=s_{ij}c_dc_e/100$, where $s_{ij}$ is the archived synergy coefficient; $t_{id}=w_i c_d$ uses an illustrative mechanism-class weight. We use the archived coefficients without refitting. The package includes the source arrays, metadata, builder and their hashes.

For the new panels, let $l_v=h_v-0.6t_v$, let $S=\sum_v x_v$, and let $S_i$ count active dose bits for compound $i$. The minimized cost, up to a constant, is
\begin{equation}
C(x)=-f(x)+P(S-2)(S-3)+P\sum_i S_i(S_i-1),
\end{equation}
where $P=3[\max_v|l_v|+\max_v\sum_w|J_{vw}|+10^{-6}]$ and $J$ is the symmetric score-interaction matrix. The second penalty therefore contributes $2P$ for each simultaneously active pair in a drug block. No mechanism-class penalty is needed for these distinct-class panels. We verify the reconstructed Ising energy against this polynomial over every binary state, up to its constant shift.

The initial probabilities use the standardized linear scores $g_v=(l_v-\bar l)/\mathrm{sd}(l)$, with the unstandardized values used if the standard deviation is below $10^{-9}$. Compute $a_v=[1+\exp(-g_v/0.5)]^{-1}$ and $c_v=\mathrm{clip}[2.5a_v/\sum_w a_w,0.001,0.999]$, then $\theta_v=2\arcsin\sqrt{c_v}$. Each training seed draws $\gamma,\beta$ independently from $N(0,0.1^2)$ and performs 60 Adam updates of the expected penalized cost. Learning rate is 0.1, moment coefficients are 0.9 and 0.99, and the numerical stabilizer is $10^{-8}$, matching the source optimizer's defaults. We retain the final iterate, with no selection by optimum probability. Every initialization, gradient, cost trajectory and final circuit is saved. The same training rule is used for all new panels. The historical hardware training trajectories are unavailable.

\section{Panel and initialization sensitivity}
Using NumPy random seed 20260905, four panels were drawn without replacement from the lexicographically ordered 35 four-compound subsets of the archived seven-compound order, before optimization. Their membership is:
\begin{itemize}
\item Panel 1: FASN, MAP2K\_1, PIK3CB\_PIK3CD, SLC16A4.
\item Panel 2: AKT\_1, FASN, MTOR\_1, SLC16A4.
\item Panel 3: FASN, MAP2K\_1, MTOR\_1, PIK3CB\_PIK3CD.
\item Panel 4: BCL2\_BCL2L1, FASN, MTOR\_1, SLC16A4.
\end{itemize}
Each panel has 12 qubits, 162 feasible assignments and one optimum. The design and source-file hashes are in the reproducibility package. All three initialization seeds per panel are included in Table~S3. The expanded CAMA-1 and second-cell-line results follow in Table~S11.

\begin{table}[h!]
\caption{Simulation results for all four panels and three seeds. Valid: ideal-QAOA feasible probability. Remaining columns give per-shot optimum probabilities; all entries are percentages. The uniform-feasible probability is $100/162$. Greedy search independently finds every panel optimum in 36, 50, 36 and 36 evaluations, respectively. All rows are noiseless simulations.}
\centering\small
\begin{tabular}{rrrrrr}
\toprule
Panel & Seed & Valid & Ideal optimum & Prior optimum & Feasible optimum\\
\midrule
\PanelRows
\bottomrule
\end{tabular}
\end{table}

The four original panels have exact optimum scores 0.327939, 0.732473, 0.532951 and 0.143720, respectively. The simulations use NumPy on Pawsey Setonix ARM CPUs. The statevector gradients and Adam updates were independently checked against PennyLane.

\section{Solution quality and repeated executions}
\begin{table}[h!]
\caption{Original six-task reference: feasible solution quality and classical search. Mean quality is $100\,\mathbb{E}[u\mid\mathcal F]$ and best gap is $100(1-\max u)$ over valid shots. Greedy gaps use the same feasible score range; the cap is 200 unique objective evaluations including partial construction. These counts do not match the full quantum computational cost.}
\centering\small
\begin{tabular}{rrrrr}
\toprule
$q$ & HW mean quality (\%) & HW best gap (\%) & Greedy gap (\%) & Eval.\\
\midrule
\QualityRows
\bottomrule
\end{tabular}
\end{table}

\begin{table}[h!]
\caption{Feasible quality for all 25 tasks. UTC date/time is in 2026. Valid: number of feasible shots; mean and gap are percentages under the normalization in Table~S4. Greedy: objective evaluations; every row's search reaches its exact optimum. Repeated problems share the same deterministic search result.}
\centering\small
\begin{tabular}{rrlrrrr}
\toprule
$q$ & $D$ & UTC date/time & Valid & Mean & Best gap & Greedy\\
\midrule
\RecoveredQualityRows
\bottomrule
\end{tabular}
\end{table}

We found identical submitted circuits in two executions at 6 qubits, three at 18, two at 20, three at 21 and two at 25. Each execution has a distinct task identifier and histogram. We report the denominator and Wilson interval for each task separately. With so few runs and no calibration records, we cannot estimate long-term drift or establish equivalence between executions. The additional programs at 6, 8, 10 and 15 qubits use different variational angles from the original series, and the two 12-qubit programs encode different instances. Those tasks remain separate comparisons. The runner records transferred parameters for the 25-, 30- and 35-qubit circuits.

\section{Hardware score distributions}
For each valid hardware output, normalize the frozen score by the full feasible score range: $u=(f-f_{\min})/(f_\star-f_{\min})$. Figure~S1 shows the empirical cumulative distribution conditional on validity, with the valid-shot denominator shown in each panel. The means and best scores summarize the observed batch.

\begin{figure}[h!]
\centering
\includegraphics[width=\textwidth]{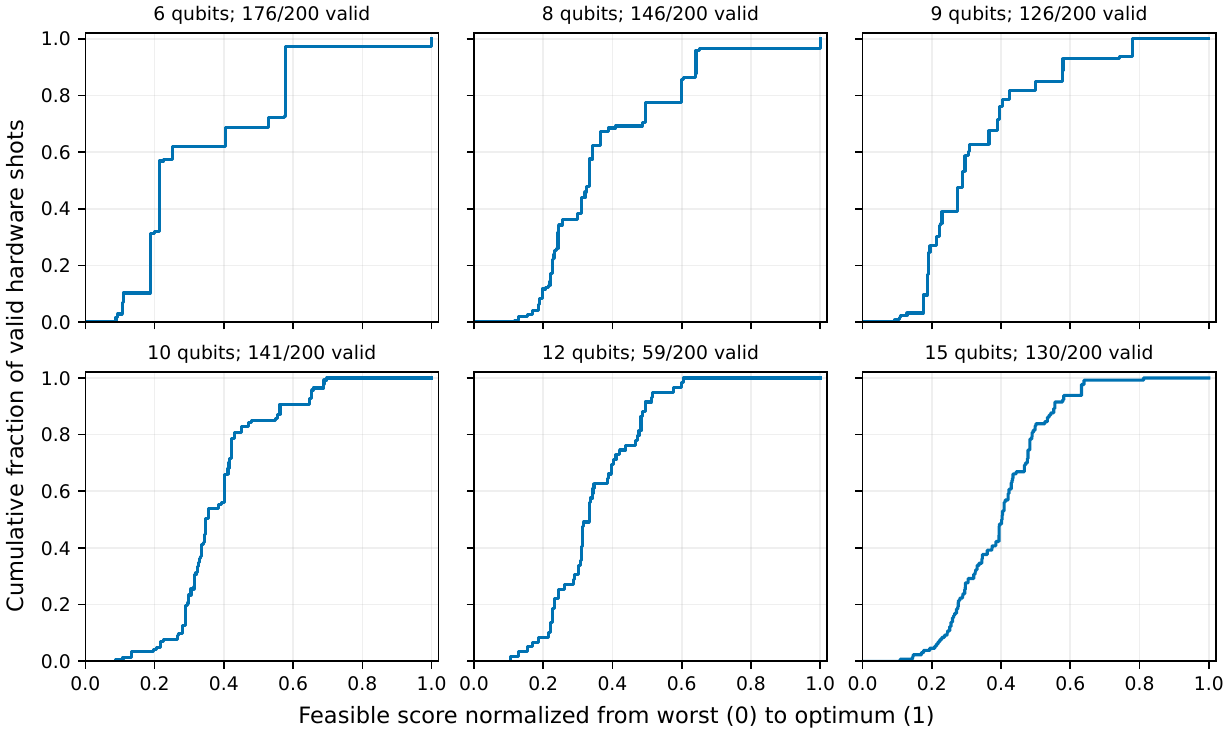}
\caption{Empirical feasible-score distributions for all six hardware batches. The horizontal axis places the worst feasible score at zero and the optimum at one; the vertical axis is the fraction of valid shots with score at or below that value. Invalid outputs are excluded from these conditional distributions and counted separately in the main feasibility analysis.}
\end{figure}

\begin{figure}[h!]
\centering
\includegraphics[width=\textwidth]{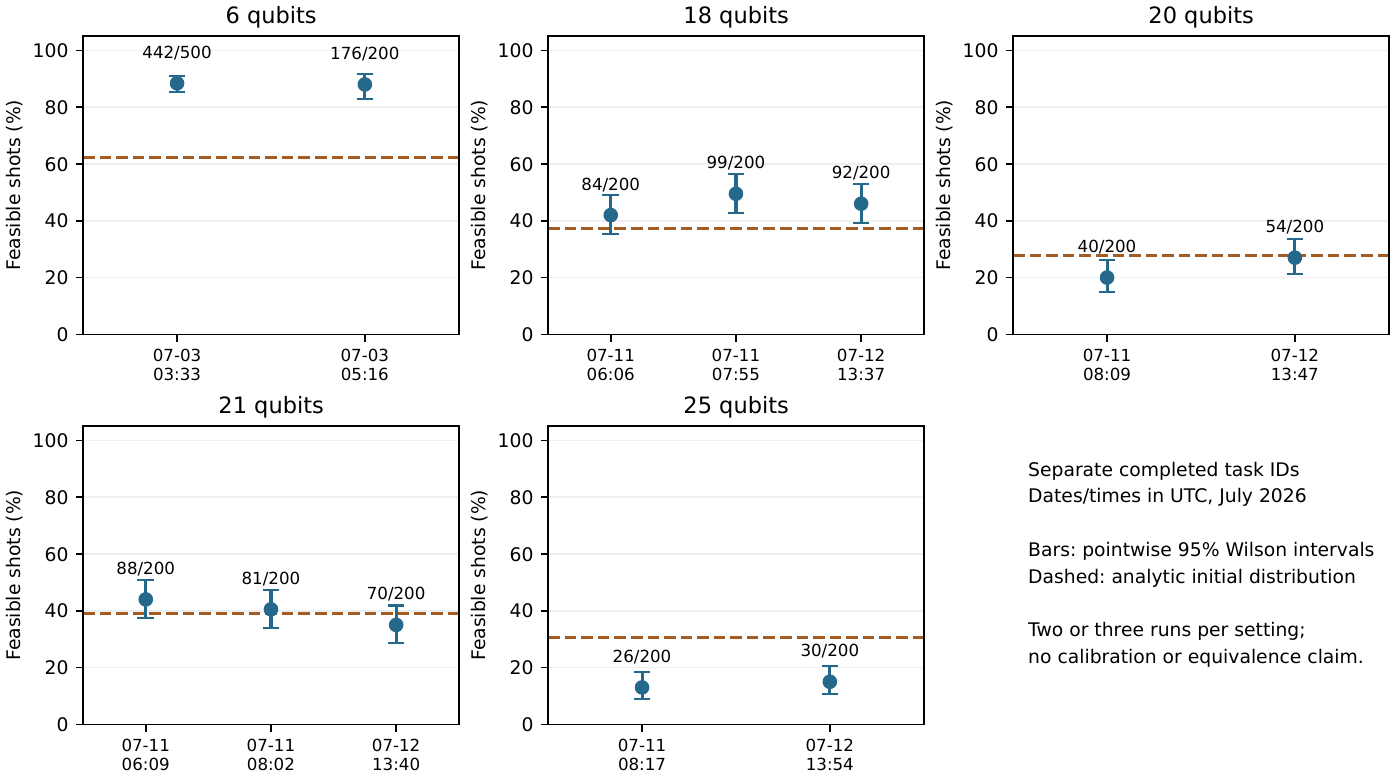}
\caption{All identical-circuit repeat groups. Markers show separate completed hardware tasks; labels give valid/total shots. Error bars are pointwise 95\% Wilson intervals, and dashed lines are analytic initial-product feasibility for that circuit. Dates and times are UTC. Each interval refers to one execution.}
\end{figure}

\clearpage
\section{Larger-circuit noiseless references}
We calculate the exact statevector of each distinct archived circuit at 18, 20, 21 and 25 qubits, retaining the original angles and coefficients. There is one circuit at each width, corresponding to three, two, three and two hardware executions. A single reference distribution is therefore used for each group. The implementation forms initial amplitudes and diagonal Ising phases in chunks of $2^{18}$ integer indices. For an index $j$, the bit on wire $v$ is $(j\mathbin{\mathrm{div}}2^{q-1-v})\bmod 2$; pairwise $Z$ eigenvalues are computed from bit parity. Local mixers update disjoint pairs of complex amplitudes in place, using bounded temporary arrays. This needs $O(2^q+C)$ working memory for chunk size $C$, rather than a $2^q\times q$ bit matrix. At 25 qubits the complex128 statevector occupies 512 MiB. This remains a full statevector calculation, with exponential storage in width.

Only feasible assignments are decoded into a small bit matrix. We sum their probabilities to obtain feasibility, sum over all tied optimal assignments for optimum probability, and probability-weight their scores to obtain conditional mean quality. The original four feasible sets contain 675, 650, 1,134 and 1,500 assignments, respectively; the two added sets contain 2,875 and 4,900. Each has one optimum. Initial-product and uniform-feasible probabilities use these same independently checked assignments. Table~S6 gives the numerically evaluated controls; Table~S7 gives each observed hardware batch. The machine-readable comparison also includes unnormalized mean scores, full-precision probabilities, task UUIDs and source hashes.

The chunked implementation reproduces all six original probability vectors within $10^{-12}$. At 18 qubits, independent simulation of the submitted gate sequence agrees within $10^{-10}$. All four final state norms differ from one by less than $10^{-10}$.

The calculations use Python 3.11.7 and NumPy 1.24.4 on Pawsey Setonix ARM CPUs. The revision adds full-statevector calculations at 30 qubits (2,875 feasible assignments) and 35 qubits (4,900 feasible assignments), using Numba 0.60.0 and complex64 amplitudes. These statevectors occupy 8 GiB and 256 GiB, respectively. We allocate four GPU shares to obtain sufficient host memory and perform the calculation on 72 CPU threads. GPU kernels are not used.

We initialize amplitudes and phases in Gray-code order within blocks of $2^{16}$ states. Each transition changes one bit, so its energy update requires only that bit's incident interactions. We recompute the starting amplitude and energy at each block boundary. Local mixers then update disjoint amplitude pairs in parallel. We save every feasible-state probability for later rescoring, without storing full statevectors in the result archive.

At 25 qubits, the maximum complex64 amplitude error relative to the original complex128 implementation is $4.14\times10^{-8}$, and the total feasible probability differs by $2.29\times10^{-10}$. The maximum relative error among feasible probabilities is $7.77\times10^{-7}$. Independent comparisons on all six original circuits pass for both precisions. State norms at 30 and 35 qubits differ from one by less than $4\times10^{-8}$. The two new calculations take 8.90 and 270.46 seconds, respectively, excluding queueing and environment setup. These are simulation runtimes, not an end-to-end comparison with quantum hardware.

\begin{table}[h!]
\caption{Noiseless and classical controls for six distinct circuits. All probability and normalized-quality entries are percentages. $|\mathcal F|$: feasible assignments; $P_{\mathcal F}$: feasible probability; $P_\star$: per-draw optimum probability; mean $u$: quality conditional on feasibility. Identical hardware repetitions share each ideal reference.}
\centering\small
\begin{tabular}{rrlrrr}
\toprule
$q$ & $|\mathcal F|$ & Distribution & $P_{\mathcal F}$ & $P_\star$ & Mean $u$\\
\midrule
\LargerControlRows
\bottomrule
\end{tabular}
\end{table}

\begin{table}[h!]
\caption{Individual larger-width hardware executions for comparison with Table~S6. Each batch has 200 shots; entries are percentages. Times are UTC in 2026. $P_\star$ here is the observed optimum fraction, not an exact probability for future executions. Pointwise feasibility intervals are in main Table~2 and Figure~1B; Table~S5 also reports best-sample gaps.}
\centering\small
\begin{tabular}{rlrrr}
\toprule
$q$ & UTC date/time & Feasible & Optimum & Mean $u$\\
\midrule
\LargerHardwareRows
\bottomrule
\end{tabular}
\end{table}

\clearpage
\clearpage
\section{Additional classical search and panel definitions}
We use simulated annealing (SA) with the same feasible dose-change, addition, deletion and compound-swap neighborhood as greedy local search. We draw the initial assignment uniformly from the feasible space by rejection sampling complete compound assignments. At each step we propose a uniformly chosen neighbor and accept an improving move, or accept a worse move with probability $\exp(\Delta f/T)$. We use 100 seeds, numbered 0 through 99, for each model and weight.

We set the initial temperature to the maximum absolute linear or pair coefficient, with a lower bound of $10^{-6}$. Over 2,000 proposals we cool geometrically to one thousandth of that temperature. We stop after 200 unique objective evaluations or 2,000 proposals, whichever comes first. Cached evaluations do not count again. The initial state's evaluation counts toward the cap. For small feasible sets the proposal limit prevents an endless attempt to obtain 200 distinct evaluations. We record the best score evaluated, and compare its trace with the enumerated optimum only after the run. Median first-hit evaluations condition on successful seeds; the success fraction reports failures separately.

Exhaustive enumeration is the exact solver: it evaluates each feasible assignment once. It supplies no search information to SA or greedy. Table~S8 reports all hardware models and the expanded panels at the archived weight. OR-Tools is not installed in the execution environment, so CP-SAT is not included.

\begingroup
\small
\setlength{\tabcolsep}{4pt}
\begin{longtable}{lrrrrrr}
\caption{Classical search at $\lambda=0.6$. H01--H14 are hardware models, C01--C35 are all CAMA-1 panels, and V1--V4 are 647-V panels. Exact: enumeration evaluations, equal to $|\mathcal F|$. Greedy: unique evaluations and optimum recovery. SA: percentage of 100 seeds reaching the optimum and median evaluations to first hit among successful seeds.}\\
\toprule
Model & $q$ & Exact & Greedy eval. & Hit & SA (\%) & First hit\\
\midrule
\endfirsthead
\toprule
Model & $q$ & Exact & Greedy eval. & Hit & SA (\%) & First hit\\
\midrule
\endhead
H01 & 6 & 20 & 21 & yes & 100 & 11 \\
H02 & 8 & 56 & 37 & yes & 100 & 20.5 \\
H03 & 9 & 54 & 33 & yes & 100 & 22.5 \\
H04 & 10 & 120 & 53 & yes & 100 & 36 \\
H05 & 12 & 162 & 57 & yes & 100 & 54 \\
H06 & 15 & 360 & 81 & yes & 98 & 87.5 \\
H07 & 18 & 675 & 105 & yes & 68 & 112.5 \\
H08 & 21 & 1134 & 129 & yes & 65 & 115 \\
H09 & 20 & 650 & 97 & yes & 65 & 115 \\
H10 & 25 & 1500 & 137 & yes & 33 & 127 \\
H11 & 12 & 220 & 69 & yes & 100 & 57 \\
H12 & 14 & 364 & 85 & yes & 95 & 77 \\
H13 & 30 & 2875 & 177 & yes & 22 & 97.5 \\
H14 & 35 & 4900 & 200 & yes & 15 & 144 \\
C01 & 12 & 162 & 36 & yes & 100 & 56 \\
C02 & 12 & 162 & 50 & yes & 100 & 56.5 \\
C03 & 12 & 162 & 50 & yes & 100 & 58 \\
C04 & 12 & 162 & 36 & yes & 100 & 61.5 \\
C05 & 12 & 162 & 50 & yes & 100 & 56 \\
C06 & 12 & 162 & 50 & yes & 100 & 49.5 \\
C07 & 12 & 162 & 36 & yes & 100 & 58.5 \\
C08 & 12 & 162 & 57 & yes & 100 & 47 \\
C09 & 12 & 162 & 50 & yes & 100 & 43.5 \\
C10 & 12 & 162 & 59 & yes & 100 & 66 \\
C11 & 12 & 162 & 50 & yes & 100 & 55.5 \\
C12 & 12 & 162 & 50 & yes & 100 & 58 \\
C13 & 12 & 162 & 36 & yes & 100 & 65 \\
C14 & 12 & 162 & 57 & yes & 100 & 50.5 \\
C15 & 12 & 162 & 50 & yes & 100 & 48.5 \\
C16 & 12 & 162 & 59 & yes & 100 & 62 \\
C17 & 12 & 162 & 57 & yes & 100 & 45 \\
C18 & 12 & 162 & 50 & yes & 100 & 43.5 \\
C19 & 12 & 162 & 59 & yes & 100 & 69.5 \\
C20 & 12 & 162 & 57 & yes & 100 & 54 \\
C21 & 12 & 162 & 36 & yes & 100 & 65 \\
C22 & 12 & 162 & 36 & yes & 100 & 53 \\
C23 & 12 & 162 & 36 & yes & 100 & 63.5 \\
C24 & 12 & 162 & 36 & yes & 100 & 45.5 \\
C25 & 12 & 162 & 36 & yes & 100 & 69.5 \\
C26 & 12 & 162 & 36 & yes & 100 & 60.5 \\
C27 & 12 & 162 & 36 & yes & 100 & 65.5 \\
C28 & 12 & 162 & 36 & yes & 100 & 59 \\
C29 & 12 & 162 & 36 & yes & 100 & 50.5 \\
C30 & 12 & 162 & 36 & yes & 100 & 38 \\
C31 & 12 & 162 & 36 & yes & 100 & 49 \\
C32 & 12 & 162 & 36 & yes & 100 & 53.5 \\
C33 & 12 & 162 & 36 & yes & 100 & 60 \\
C34 & 12 & 162 & 36 & yes & 100 & 52 \\
C35 & 12 & 162 & 36 & yes & 100 & 54 \\
V1 & 12 & 162 & 46 & yes & 100 & 58 \\
V2 & 12 & 162 & 46 & yes & 100 & 48.5 \\
V3 & 12 & 162 & 55 & yes & 100 & 58.5 \\
V4 & 12 & 162 & 50 & no & 100 & 84 \\
\bottomrule
\end{longtable}
\endgroup

Hardware model labels follow the original six instances first, then the additional distinct models. Their compound counts, dose counts and feasible-set sizes identify the encoding; Table~S9 additionally identifies every distinct circuit. Panel membership is given with Table~S11.

\clearpage
\section{Score-weight sensitivity}
We retain $\lambda=0.6$ as the archived model value and rescore each feasible assignment at $\lambda\in\{0.3,0.45,0.6,0.75,0.9\}$. All archived circuit coefficients, initialization angles, variational angles and hardware histograms remain fixed. Only the score and the set of globally optimal feasible assignments change. We sum each fixed distribution over that new optimum set, including ties. For the original four 12-qubit panels we instead rebuild the penalty cost and product prior at each weight and retrain all three seeds with 60 Adam steps and learning rate 0.1. Table~S9 distinguishes those retrained panels (P1--P4) from the fixed circuits (Q01--Q18).

Each entry below is a per-draw optimum probability, in percent. Hardware values are observed fractions, with repeated tasks separated by slashes in chronological order. A dash indicates an unavailable noiseless reference. Greedy reports whether it reaches the re-enumerated optimum; SA gives the percentage of successful seeds. For retrained panels the ideal values follow seed order 0, 1 and 2. The machine-readable results include exact optimum scores, all tied optimum bitstrings, individual hardware task identifiers and every SA seed.

\begingroup
\small
\setlength{\tabcolsep}{4pt}
\begin{longtable}{rrrrlrrr}
\caption{Weight sensitivity for every archived circuit and the original four retrained panels. Prior, Ideal, Feasible and HW are optimum probabilities (percent). Hit: greedy optimum recovery. SA: successful seeds (percent); Eval.: median first-hit evaluations among those seeds.}\\
\toprule
$\lambda$ & Prior & Ideal & Feasible & HW & Hit & SA & Eval.\\
\midrule
\endfirsthead
\toprule
$\lambda$ & Prior & Ideal & Feasible & HW & Hit & SA & Eval.\\
\midrule
\endhead
\multicolumn{8}{l}{Q01: H01, 6 qubits, 20 feasible assignments} \\
0.3 & 0.9404 & 1.5739 & 5.0000 & 2.50 & yes & 100 & 7.5 \\
0.45 & 0.9404 & 1.5739 & 5.0000 & 2.50 & yes & 100 & 8 \\
0.6 & 0.9404 & 1.5739 & 5.0000 & 2.50 & yes & 100 & 11 \\
0.75 & 0.9404 & 1.5739 & 5.0000 & 2.50 & yes & 100 & 13 \\
0.9 & 0.9404 & 1.5739 & 5.0000 & 2.50 & no & 100 & 11 \\
\multicolumn{8}{l}{Q02: H01, 6 qubits, 20 feasible assignments} \\
0.3 & 0.9404 & 1.6009 & 5.0000 & 1.20/2.50 & yes & 100 & 7.5 \\
0.45 & 0.9404 & 1.6009 & 5.0000 & 1.20/2.50 & yes & 100 & 8 \\
0.6 & 0.9404 & 1.6009 & 5.0000 & 1.20/2.50 & yes & 100 & 11 \\
0.75 & 0.9404 & 1.6009 & 5.0000 & 1.20/2.50 & yes & 100 & 13 \\
0.9 & 0.9404 & 1.6009 & 5.0000 & 1.20/2.50 & no & 100 & 11 \\
\multicolumn{8}{l}{Q03: H02, 8 qubits, 56 feasible assignments} \\
0.3 & 1.3671 & 2.3948 & 1.7857 & 2.50 & yes & 100 & 17 \\
0.45 & 1.3671 & 2.3948 & 1.7857 & 2.50 & yes & 100 & 19.5 \\
0.6 & 1.3671 & 2.3948 & 1.7857 & 2.50 & yes & 100 & 20.5 \\
0.75 & 1.3671 & 2.3948 & 1.7857 & 2.50 & yes & 100 & 24 \\
0.9 & 1.3671 & 2.3948 & 1.7857 & 2.50 & no & 100 & 29 \\
\multicolumn{8}{l}{Q04: H02, 8 qubits, 56 feasible assignments} \\
0.3 & 1.3671 & 2.3900 & 1.7857 & 2.50 & yes & 100 & 17 \\
0.45 & 1.3671 & 2.3900 & 1.7857 & 2.50 & yes & 100 & 19.5 \\
0.6 & 1.3671 & 2.3900 & 1.7857 & 2.50 & yes & 100 & 20.5 \\
0.75 & 1.3671 & 2.3900 & 1.7857 & 2.50 & yes & 100 & 24 \\
0.9 & 1.3671 & 2.3900 & 1.7857 & 2.50 & no & 100 & 29 \\
\multicolumn{8}{l}{Q05: H03, 9 qubits, 54 feasible assignments} \\
0.3 & 0.0550 & 0.0928 & 1.8519 & 0.00 & yes & 100 & 19 \\
0.45 & 0.0550 & 0.0928 & 1.8519 & 0.00 & yes & 100 & 22 \\
0.6 & 0.0550 & 0.0928 & 1.8519 & 0.00 & yes & 100 & 22.5 \\
0.75 & 0.0550 & 0.0928 & 1.8519 & 0.00 & yes & 100 & 28 \\
0.9 & 0.0550 & 0.0928 & 1.8519 & 0.00 & yes & 100 & 28.5 \\
\multicolumn{8}{l}{Q06: H04, 10 qubits, 120 feasible assignments} \\
0.3 & 0.8128 & 1.4110 & 0.8333 & 2.00 & yes & 100 & 27 \\
0.45 & 0.8128 & 1.4110 & 0.8333 & 2.00 & yes & 100 & 32.5 \\
0.6 & 0.8128 & 1.4110 & 0.8333 & 2.00 & yes & 100 & 36 \\
0.75 & 0.8128 & 1.4110 & 0.8333 & 2.00 & yes & 100 & 40 \\
0.9 & 0.8128 & 1.4110 & 0.8333 & 2.00 & no & 100 & 43 \\
\multicolumn{8}{l}{Q07: H04, 10 qubits, 120 feasible assignments} \\
0.3 & 0.8128 & 1.3987 & 0.8333 & 0.00 & yes & 100 & 27 \\
0.45 & 0.8128 & 1.3987 & 0.8333 & 0.00 & yes & 100 & 32.5 \\
0.6 & 0.8128 & 1.3987 & 0.8333 & 0.00 & yes & 100 & 36 \\
0.75 & 0.8128 & 1.3987 & 0.8333 & 0.00 & yes & 100 & 40 \\
0.9 & 0.8128 & 1.3987 & 0.8333 & 0.00 & no & 100 & 43 \\
\multicolumn{8}{l}{Q08: H11, 12 qubits, 220 feasible assignments} \\
0.3 & 0.4592 & 0.7827 & 0.4545 & 2.00 & yes & 100 & 52 \\
0.45 & 0.4592 & 0.7827 & 0.4545 & 2.00 & yes & 100 & 52.5 \\
0.6 & 0.4592 & 0.7827 & 0.4545 & 2.00 & yes & 100 & 57 \\
0.75 & 0.4592 & 0.7827 & 0.4545 & 2.00 & yes & 100 & 79 \\
0.9 & 0.4592 & 0.7827 & 0.4545 & 2.00 & no & 100 & 77.5 \\
\multicolumn{8}{l}{Q09: H05, 12 qubits, 162 feasible assignments} \\
0.3 & 0.1657 & 0.1917 & 0.6173 & 0.00 & yes & 100 & 44.5 \\
0.45 & 0.1657 & 0.1917 & 0.6173 & 0.00 & yes & 100 & 46 \\
0.6 & 0.1657 & 0.1917 & 0.6173 & 0.00 & yes & 100 & 54 \\
0.75 & 0.1657 & 0.1917 & 0.6173 & 0.00 & yes & 100 & 59 \\
0.9 & 0.1657 & 0.1917 & 0.6173 & 0.00 & yes & 100 & 62 \\
\multicolumn{8}{l}{Q10: H12, 14 qubits, 364 feasible assignments} \\
0.3 & 0.2963 & 0.5040 & 0.2747 & 1.50 & yes & 100 & 72 \\
0.45 & 0.2963 & 0.5040 & 0.2747 & 1.50 & yes & 99 & 67 \\
0.6 & 0.2963 & 0.5040 & 0.2747 & 1.50 & yes & 95 & 77 \\
0.75 & 0.2963 & 0.5040 & 0.2747 & 1.50 & yes & 97 & 96 \\
0.9 & 0.2963 & 0.5040 & 0.2747 & 1.50 & no & 77 & 109 \\
\multicolumn{8}{l}{Q11: H06, 15 qubits, 360 feasible assignments} \\
0.3 & 0.1438 & 0.2469 & 0.2778 & 0.00 & yes & 98 & 58 \\
0.45 & 0.1438 & 0.2469 & 0.2778 & 0.00 & yes & 99 & 66 \\
0.6 & 0.1438 & 0.2469 & 0.2778 & 0.00 & yes & 98 & 87.5 \\
0.75 & 0.1438 & 0.2469 & 0.2778 & 0.00 & yes & 85 & 86 \\
0.9 & 0.1438 & 0.2469 & 0.2778 & 0.00 & yes & 75 & 110 \\
\multicolumn{8}{l}{Q12: H06, 15 qubits, 360 feasible assignments} \\
0.3 & 0.1438 & 0.2456 & 0.2778 & 0.00 & yes & 98 & 58 \\
0.45 & 0.1438 & 0.2456 & 0.2778 & 0.00 & yes & 99 & 66 \\
0.6 & 0.1438 & 0.2456 & 0.2778 & 0.00 & yes & 98 & 87.5 \\
0.75 & 0.1438 & 0.2456 & 0.2778 & 0.00 & yes & 85 & 86 \\
0.9 & 0.1438 & 0.2456 & 0.2778 & 0.00 & yes & 75 & 110 \\
\multicolumn{8}{l}{Q13: H07, 18 qubits, 675 feasible assignments} \\
0.3 & 0.0985 & 0.1666 & 0.1481 & 0.00/0.50/0.50 & yes & 88 & 93.5 \\
0.45 & 0.0985 & 0.1666 & 0.1481 & 0.00/0.50/0.50 & yes & 76 & 73 \\
0.6 & 0.0985 & 0.1666 & 0.1481 & 0.00/0.50/0.50 & yes & 68 & 112.5 \\
0.75 & 0.0985 & 0.1666 & 0.1481 & 0.00/0.50/0.50 & yes & 56 & 116.5 \\
0.9 & 0.0985 & 0.1666 & 0.1481 & 0.00/0.50/0.50 & yes & 39 & 81 \\
\multicolumn{8}{l}{Q14: H09, 20 qubits, 650 feasible assignments} \\
0.3 & 0.0212 & 0.0239 & 0.1538 & 0.00/0.00 & yes & 79 & 87 \\
0.45 & 0.0212 & 0.0239 & 0.1538 & 0.00/0.00 & yes & 84 & 104.5 \\
0.6 & 0.0212 & 0.0239 & 0.1538 & 0.00/0.00 & yes & 65 & 115 \\
0.75 & 0.0212 & 0.0239 & 0.1538 & 0.00/0.00 & yes & 59 & 118 \\
0.9 & 0.0212 & 0.0239 & 0.1538 & 0.00/0.00 & yes & 51 & 123 \\
\multicolumn{8}{l}{Q15: H08, 21 qubits, 1,134 feasible assignments} \\
0.3 & 0.0670 & 0.1111 & 0.0882 & 0.50/0.00/0.00 & yes & 68 & 121 \\
0.45 & 0.0670 & 0.1111 & 0.0882 & 0.50/0.00/0.00 & yes & 75 & 123 \\
0.6 & 0.0670 & 0.1111 & 0.0882 & 0.50/0.00/0.00 & yes & 65 & 115 \\
0.75 & 0.0670 & 0.1111 & 0.0882 & 0.50/0.00/0.00 & yes & 42 & 94 \\
0.9 & 0.0670 & 0.1111 & 0.0882 & 0.50/0.00/0.00 & yes & 31 & 94 \\
\multicolumn{8}{l}{Q16: H10, 25 qubits, 1,500 feasible assignments} \\
0.3 & 0.0222 & 0.0242 & 0.0667 & 0.00/0.00 & yes & 51 & 121 \\
0.45 & 0.0222 & 0.0242 & 0.0667 & 0.00/0.00 & yes & 43 & 151 \\
0.6 & 0.0222 & 0.0242 & 0.0667 & 0.00/0.00 & yes & 33 & 127 \\
0.75 & 0.0222 & 0.0242 & 0.0667 & 0.00/0.00 & yes & 24 & 131 \\
0.9 & 0.0222 & 0.0242 & 0.0667 & 0.00/0.00 & yes & 18 & 147.5 \\
\multicolumn{8}{l}{Q17: H13, 30 qubits, 2,875 feasible assignments} \\
0.3 & 0.0172 & 0.0174 & 0.0348 & 0.00 & yes & 39 & 83 \\
0.45 & 0.0172 & 0.0174 & 0.0348 & 0.00 & yes & 31 & 118 \\
0.6 & 0.0172 & 0.0174 & 0.0348 & 0.00 & yes & 22 & 97.5 \\
0.75 & 0.0172 & 0.0174 & 0.0348 & 0.00 & no & 10 & 95 \\
0.9 & 0.0172 & 0.0174 & 0.0348 & 0.00 & no & 13 & 108 \\
\multicolumn{8}{l}{Q18: H14, 35 qubits, 4,900 feasible assignments} \\
0.3 & 0.0122 & 0.0120 & 0.0204 & 0.00 & yes & 25 & 124 \\
0.45 & 0.0122 & 0.0120 & 0.0204 & 0.00 & yes & 17 & 132 \\
0.6 & 0.0122 & 0.0120 & 0.0204 & 0.00 & yes & 15 & 144 \\
0.75 & 0.0122 & 0.0120 & 0.0204 & 0.00 & no & 9 & 83 \\
0.9 & 0.0122 & 0.0120 & 0.0204 & 0.00 & no & 6 & 158.5 \\
\multicolumn{8}{l}{P1: retrained 12-qubit panel, 162 feasible assignments} \\
0.3 & 0.0438 & 0.07/0.08/0.07 & 0.6173 & -- & no & 100 & 62 \\
0.45 & 1.1547 & 2.56/2.48/2.62 & 0.6173 & -- & yes & 100 & 68 \\
0.6 & 1.2651 & 2.68/2.23/2.75 & 0.6173 & -- & yes & 100 & 60 \\
0.75 & 1.3438 & 2.91/2.75/2.90 & 0.6173 & -- & yes & 100 & 57 \\
0.9 & 1.3615 & 2.81/2.59/2.90 & 0.6173 & -- & yes & 100 & 60.5 \\
\multicolumn{8}{l}{P2: retrained 12-qubit panel, 162 feasible assignments} \\
0.3 & 3.0816 & 6.77/6.85/6.64 & 0.6173 & -- & yes & 100 & 44.5 \\
0.45 & 2.3851 & 5.43/5.12/5.40 & 0.6173 & -- & yes & 100 & 51.5 \\
0.6 & 1.4854 & 3.36/3.32/3.43 & 0.6173 & -- & yes & 100 & 48.5 \\
0.75 & 0.5827 & 1.39/1.46/1.38 & 0.6173 & -- & no & 100 & 54.5 \\
0.9 & 1.4192 & 3.07/2.88/2.95 & 0.6173 & -- & yes & 100 & 50.5 \\
\multicolumn{8}{l}{P3: retrained 12-qubit panel, 162 feasible assignments} \\
0.3 & 0.1872 & 0.32/0.33/0.32 & 0.6173 & -- & yes & 100 & 46 \\
0.45 & 0.3624 & 0.61/0.62/0.61 & 0.6173 & -- & yes & 100 & 64.5 \\
0.6 & 1.4361 & 3.28/3.25/3.22 & 0.6173 & -- & yes & 100 & 49 \\
0.75 & 1.4490 & 3.06/1.33/3.09 & 0.6173 & -- & yes & 100 & 62.5 \\
0.9 & 1.3731 & 2.86/2.74/2.90 & 0.6173 & -- & yes & 100 & 59 \\
\multicolumn{8}{l}{P4: retrained 12-qubit panel, 162 feasible assignments} \\
0.3 & 1.2864 & 2.86/2.83/2.84 & 0.6173 & -- & yes & 100 & 69 \\
0.45 & 1.3741 & 3.07/3.04/3.00 & 0.6173 & -- & yes & 100 & 50 \\
0.6 & 1.2954 & 2.89/2.81/2.87 & 0.6173 & -- & yes & 100 & 69.5 \\
0.75 & 1.2801 & 2.77/2.71/2.75 & 0.6173 & -- & yes & 100 & 54 \\
0.9 & 1.2979 & 2.74/2.69/2.75 & 0.6173 & -- & yes & 100 & 59 \\
\bottomrule
\end{longtable}
\endgroup

Circuit identifiers map to the archived task inventory as follows. Repeated tasks share a circuit identifier.
\begin{itemize}

\item Q01: \texttt{c8ea4f7a-a64e-49d2-a7f8-f1ac1eff2136}.

\item Q02: \texttt{24d41998-5edf-44bf-93a4-60fca948ae8c}\newline \texttt{98cb6dee-1d5d-4aec-9122-6a1c390a2608}.

\item Q03: \texttt{e1a915b0-ff23-4769-8289-bfb7fcb0f74b}.

\item Q04: \texttt{4e307518-f350-4843-98ab-06f91315c190}.

\item Q05: \texttt{e88a4bac-5f06-4b2c-bcc8-69a20d23e527}.

\item Q06: \texttt{fad4ee6f-3815-4d92-af93-dbd2c19ec6a4}.

\item Q07: \texttt{9fcd57c3-97a5-4070-929d-d4ba5b7284f9}.

\item Q08: \texttt{48d52888-7f37-4931-a135-8f831f1a208e}.

\item Q09: \texttt{5cecf3e8-853e-448d-b296-d8bd2ef9f5e0}.

\item Q10: \texttt{1ebf3086-bb62-48fc-9bdb-0df1dedd62cf}.

\item Q11: \texttt{4d659093-99b5-4d58-9063-cd20724f7a98}.

\item Q12: \texttt{566d8e0f-b67b-4786-86fa-3b112aaf15bd}.

\item Q13: \texttt{4b60729f-56e1-45d5-a2d6-fec25d56c84c}\newline \texttt{b3fd2bd7-1b51-48e1-9c63-cbe7d55e1120}\newline \texttt{014e4aba-2301-4b48-87ba-28c46e97e84f}.

\item Q14: \texttt{abdccbd0-9b84-4ba1-b5a7-91c6e0403be6}\newline \texttt{b4079656-2649-4163-9025-133821ff1bdd}.

\item Q15: \texttt{fa0c9bef-c035-4232-a77b-75970696b227}\newline \texttt{30a44a65-2124-4664-9720-7d3d53e7498e}\newline \texttt{44849fe4-b8c2-4128-8792-ae5ef06009eb}.

\item Q16: \texttt{e69c31e1-b34e-4869-bae9-217de933cb7d}\newline \texttt{b1386480-d348-4a01-ae1a-dbecf6d2a934}.

\item Q17: \texttt{bb016dac-cabb-4e80-b568-0c0818917ab6}.

\item Q18: \texttt{cfdf6f35-0141-46b3-a059-7f9e9fc85b9c}.

\end{itemize}

\clearpage
\section{Feasibility-preserving Grover mixer}
We construct $|\mathcal F\rangle$ directly as the uniform vector over enumerated feasible assignments. One layer applies $\exp(i\gamma f)$ followed by $I-(1-e^{-i\beta})|\mathcal F\rangle\langle\mathcal F|$. Thus the optimized energy is $-f$, with no penalty term. We train both angles from the same three random seeds, using 60 Adam steps with learning rate 0.1, and retain every final iterate. Analytic derivatives agree with centered finite differences, and a dense projector calculation independently checks the mixer. Every state remains normalized and entirely feasible.

The construction follows B{\"a}rtschi and Eidenbenz (2020), cited in the main paper. Direct feasible-vector preparation is a simulation device and does not establish a low-cost hardware preparation circuit. The comparison changes initialization and training as well as the mixer: the original penalty circuits retain their archived parameters, whereas both simulated panel methods are trained with the stated settings.

\begingroup
\small
\setlength{\tabcolsep}{4pt}
\begin{longtable}{lrrrrr}
\caption{One-layer Grover-mixer comparison for the six reference instances and four original panels. All probabilities are percentages. Grover feasibility is exactly 100 percent. Penalty: warm-start penalty-QAOA optimum probability. Greedy reaches every optimum, with the listed evaluation count.}\\
\toprule
Model & Seed & Grover & Penalty & Feasible & Greedy eval.\\
\midrule
\endfirsthead
\toprule
Model & Seed & Grover & Penalty & Feasible & Greedy eval.\\
\midrule
\endhead
H01 & 0 & 27.2214 & 1.6009 & 5.0000 & 21 \\
H01 & 1 & 27.1291 & 1.6009 & 5.0000 & 21 \\
H01 & 2 & 27.2416 & 1.6009 & 5.0000 & 21 \\
H02 & 0 & 9.4232 & 2.3900 & 1.7857 & 37 \\
H02 & 1 & 9.3654 & 2.3900 & 1.7857 & 37 \\
H02 & 2 & 9.4298 & 2.3900 & 1.7857 & 37 \\
H03 & 0 & 9.9900 & 0.0928 & 1.8519 & 33 \\
H03 & 1 & 9.9935 & 0.0928 & 1.8519 & 33 \\
H03 & 2 & 9.9820 & 0.0928 & 1.8519 & 33 \\
H04 & 0 & 3.9305 & 1.3987 & 0.8333 & 53 \\
H04 & 1 & 3.9296 & 1.3987 & 0.8333 & 53 \\
H04 & 2 & 3.9265 & 1.3987 & 0.8333 & 53 \\
H05 & 0 & 3.0192 & 0.1917 & 0.6173 & 57 \\
H05 & 1 & 3.0452 & 0.1917 & 0.6173 & 57 \\
H05 & 2 & 3.0096 & 0.1917 & 0.6173 & 57 \\
H06 & 0 & 1.1758 & 0.2469 & 0.2778 & 81 \\
H06 & 1 & 1.2041 & 0.2469 & 0.2778 & 81 \\
H06 & 2 & 1.1683 & 0.2469 & 0.2778 & 81 \\
P1 & 0 & 1.5277 & 2.6819 & 0.6173 & 36 \\
P1 & 1 & 1.5411 & 2.2299 & 0.6173 & 36 \\
P1 & 2 & 1.5259 & 2.7522 & 0.6173 & 36 \\
P2 & 0 & 2.9140 & 3.3647 & 0.6173 & 50 \\
P2 & 1 & 2.9422 & 3.3187 & 0.6173 & 50 \\
P2 & 2 & 2.9107 & 3.4274 & 0.6173 & 50 \\
P3 & 0 & 1.8449 & 3.2790 & 0.6173 & 36 \\
P3 & 1 & 1.8626 & 3.2483 & 0.6173 & 36 \\
P3 & 2 & 1.8422 & 3.2188 & 0.6173 & 36 \\
P4 & 0 & 1.7846 & 2.8943 & 0.6173 & 36 \\
P4 & 1 & 1.7842 & 2.8090 & 0.6173 & 36 \\
P4 & 2 & 1.7866 & 2.8654 & 0.6173 & 36 \\
\bottomrule
\end{longtable}
\endgroup

\clearpage
\section{All CAMA-1 panels and a second cell line}
We enumerate all 35 four-compound subsets of the seven-compound CAMA-1 model. For a second cell line, we inspect QA-passing DREAM pair measurements in alphabetical cell-line order, excluding CAMA-1. We choose the first line with a fully measured five-compound clique having distinct mechanism labels and finite fitted Hill parameters. Within a cell line, compounds are searched lexicographically. This gives 647-V and the compounds AKT, ATM, ATR\_4, CHEK1 and Cisplatin. We sample four of its five four-compound subsets without replacement using seed 20260905, before optimization.

We use the original coefficient-construction rule: median Hill parameters from within-clique observations, mean recorded pair synergy divided by 100 and multiplied by dose fractions, and the same class-weighted dose proxy. Mechanism classes absent from the archived weight dictionary use its existing default weight of 1.0. Every panel uses dose fractions $(0.1,0.32,1)$, 12 qubits and 162 feasible assignments. Table~S11 reports all three training seeds. Figure~S3 shows their optimum probabilities without selecting the best seed.

\begingroup
\small
\setlength{\tabcolsep}{4pt}
\begin{longtable}{lrrrrrr}
\caption{Expanded panel results at $\lambda=0.6$. Q0--Q2: ideal penalty-QAOA optimum probabilities for training seeds 0--2; Prior and Feasible: corresponding sampling controls. All probabilities are percentages. Hit: greedy optimum recovery. Every SA seed reaches each panel optimum (Table~S8).}\\
\toprule
Panel & Q0 & Q1 & Q2 & Prior & Feasible & Hit\\
\midrule
\endfirsthead
\toprule
Panel & Q0 & Q1 & Q2 & Prior & Feasible & Hit\\
\midrule
\endhead
C01 & 2.7201 & 2.6732 & 2.6643 & 1.2093 & 0.6173 & yes \\
C02 & 3.0407 & 1.9400 & 3.1236 & 1.3703 & 0.6173 & yes \\
C03 & 1.9339 & 1.2859 & 1.9579 & 0.7976 & 0.6173 & yes \\
C04 & 2.8402 & 2.7449 & 2.7772 & 1.2779 & 0.6173 & yes \\
C05 & 2.9004 & 2.0200 & 2.9279 & 1.2525 & 0.6173 & yes \\
C06 & 0.8662 & 1.4319 & 2.2637 & 0.8631 & 0.6173 & yes \\
C07 & 2.5968 & 1.1360 & 2.5735 & 1.2025 & 0.6173 & yes \\
C08 & 0.3154 & 0.0481 & 0.0596 & 0.1813 & 0.6173 & yes \\
C09 & 3.1359 & 1.2523 & 3.1809 & 1.3989 & 0.6173 & yes \\
C10 & 0.0340 & 0.0086 & 0.0356 & 0.0200 & 0.6173 & yes \\
C11 & 2.6345 & 1.7352 & 2.7382 & 1.1920 & 0.6173 & yes \\
C12 & 0.7304 & 1.1637 & 1.8832 & 0.7686 & 0.6173 & yes \\
C13 & 2.6084 & 2.5187 & 2.5084 & 1.1635 & 0.6173 & yes \\
C14 & 0.1652 & 0.0321 & 0.2442 & 0.1473 & 0.6173 & yes \\
C15 & 3.3647 & 3.3187 & 3.4274 & 1.4854 & 0.6173 & yes \\
C16 & 0.0090 & 0.0033 & 0.0126 & 0.0080 & 0.6173 & yes \\
C17 & 0.3170 & 0.3146 & 0.0769 & 0.1852 & 0.6173 & yes \\
C18 & 2.8396 & 1.8293 & 2.7896 & 1.2364 & 0.6173 & yes \\
C19 & 0.0521 & 0.0155 & 0.0376 & 0.0343 & 0.6173 & yes \\
C20 & 0.1917 & 0.0362 & 0.2817 & 0.1657 & 0.6173 & yes \\
C21 & 2.6873 & 2.6428 & 2.7050 & 1.2276 & 0.6173 & yes \\
C22 & 2.8960 & 2.8696 & 2.9146 & 1.3126 & 0.6173 & yes \\
C23 & 2.6387 & 2.5725 & 2.6672 & 1.2712 & 0.6173 & yes \\
C24 & 3.5930 & 3.8332 & 3.6667 & 1.6352 & 0.6173 & yes \\
C25 & 2.8943 & 2.8090 & 2.8654 & 1.2954 & 0.6173 & yes \\
C26 & 2.8975 & 2.5269 & 2.9212 & 1.3560 & 0.6173 & yes \\
C27 & 3.2734 & 3.4699 & 3.3335 & 1.4757 & 0.6173 & yes \\
C28 & 2.6831 & 2.5605 & 2.6821 & 1.2185 & 0.6173 & yes \\
C29 & 2.7682 & 2.3319 & 2.8009 & 1.2914 & 0.6173 & yes \\
C30 & 3.6023 & 3.3816 & 3.7725 & 1.6526 & 0.6173 & yes \\
C31 & 3.2790 & 3.2483 & 3.2188 & 1.4361 & 0.6173 & yes \\
C32 & 2.5639 & 2.5042 & 2.5892 & 1.1820 & 0.6173 & yes \\
C33 & 2.6819 & 2.2299 & 2.7522 & 1.2651 & 0.6173 & yes \\
C34 & 3.9957 & 3.4488 & 3.8619 & 1.7604 & 0.6173 & yes \\
C35 & 3.1989 & 2.6416 & 3.3414 & 1.4724 & 0.6173 & yes \\
V1 & 0.0097 & 0.0098 & 0.0060 & 0.0060 & 0.6173 & yes \\
V2 & 0.2359 & 0.0921 & 0.2307 & 0.1382 & 0.6173 & yes \\
V3 & 0.0016 & 0.0014 & 0.0017 & 0.0010 & 0.6173 & yes \\
V4 & 0.0039 & 0.0016 & 0.0020 & 0.0023 & 0.6173 & no \\
\bottomrule
\end{longtable}
\endgroup

\begin{itemize}

\item C01: AKT\_1, BCL2\_BCL2L1, FASN, MAP2K\_1.

\item C02: AKT\_1, BCL2\_BCL2L1, FASN, MTOR\_1.

\item C03: AKT\_1, BCL2\_BCL2L1, FASN, PIK3CB\_PIK3CD.

\item C04: AKT\_1, BCL2\_BCL2L1, FASN, SLC16A4.

\item C05: AKT\_1, BCL2\_BCL2L1, MAP2K\_1, MTOR\_1.

\item C06: AKT\_1, BCL2\_BCL2L1, MAP2K\_1, PIK3CB\_PIK3CD.

\item C07: AKT\_1, BCL2\_BCL2L1, MAP2K\_1, SLC16A4.

\item C08: AKT\_1, BCL2\_BCL2L1, MTOR\_1, PIK3CB\_PIK3CD.

\item C09: AKT\_1, BCL2\_BCL2L1, MTOR\_1, SLC16A4.

\item C10: AKT\_1, BCL2\_BCL2L1, PIK3CB\_PIK3CD, SLC16A4.

\item C11: AKT\_1, FASN, MAP2K\_1, MTOR\_1.

\item C12: AKT\_1, FASN, MAP2K\_1, PIK3CB\_PIK3CD.

\item C13: AKT\_1, FASN, MAP2K\_1, SLC16A4.

\item C14: AKT\_1, FASN, MTOR\_1, PIK3CB\_PIK3CD.

\item C15: AKT\_1, FASN, MTOR\_1, SLC16A4.

\item C16: AKT\_1, FASN, PIK3CB\_PIK3CD, SLC16A4.

\item C17: AKT\_1, MAP2K\_1, MTOR\_1, PIK3CB\_PIK3CD.

\item C18: AKT\_1, MAP2K\_1, MTOR\_1, SLC16A4.

\item C19: AKT\_1, MAP2K\_1, PIK3CB\_PIK3CD, SLC16A4.

\item C20: AKT\_1, MTOR\_1, PIK3CB\_PIK3CD, SLC16A4.

\item C21: BCL2\_BCL2L1, FASN, MAP2K\_1, MTOR\_1.

\item C22: BCL2\_BCL2L1, FASN, MAP2K\_1, PIK3CB\_PIK3CD.

\item C23: BCL2\_BCL2L1, FASN, MAP2K\_1, SLC16A4.

\item C24: BCL2\_BCL2L1, FASN, MTOR\_1, PIK3CB\_PIK3CD.

\item C25: BCL2\_BCL2L1, FASN, MTOR\_1, SLC16A4.

\item C26: BCL2\_BCL2L1, FASN, PIK3CB\_PIK3CD, SLC16A4.

\item C27: BCL2\_BCL2L1, MAP2K\_1, MTOR\_1, PIK3CB\_PIK3CD.

\item C28: BCL2\_BCL2L1, MAP2K\_1, MTOR\_1, SLC16A4.

\item C29: BCL2\_BCL2L1, MAP2K\_1, PIK3CB\_PIK3CD, SLC16A4.

\item C30: BCL2\_BCL2L1, MTOR\_1, PIK3CB\_PIK3CD, SLC16A4.

\item C31: FASN, MAP2K\_1, MTOR\_1, PIK3CB\_PIK3CD.

\item C32: FASN, MAP2K\_1, MTOR\_1, SLC16A4.

\item C33: FASN, MAP2K\_1, PIK3CB\_PIK3CD, SLC16A4.

\item C34: FASN, MTOR\_1, PIK3CB\_PIK3CD, SLC16A4.

\item C35: MAP2K\_1, MTOR\_1, PIK3CB\_PIK3CD, SLC16A4.

\item V1: ATM, ATR\_4, CHEK1, Cisplatin.

\item V2: AKT, ATM, CHEK1, Cisplatin.

\item V3: AKT, ATR\_4, CHEK1, Cisplatin.

\item V4: AKT, ATM, ATR\_4, Cisplatin.

\end{itemize}

\begin{figure}[h!]
\centering
\includegraphics[width=\textwidth]{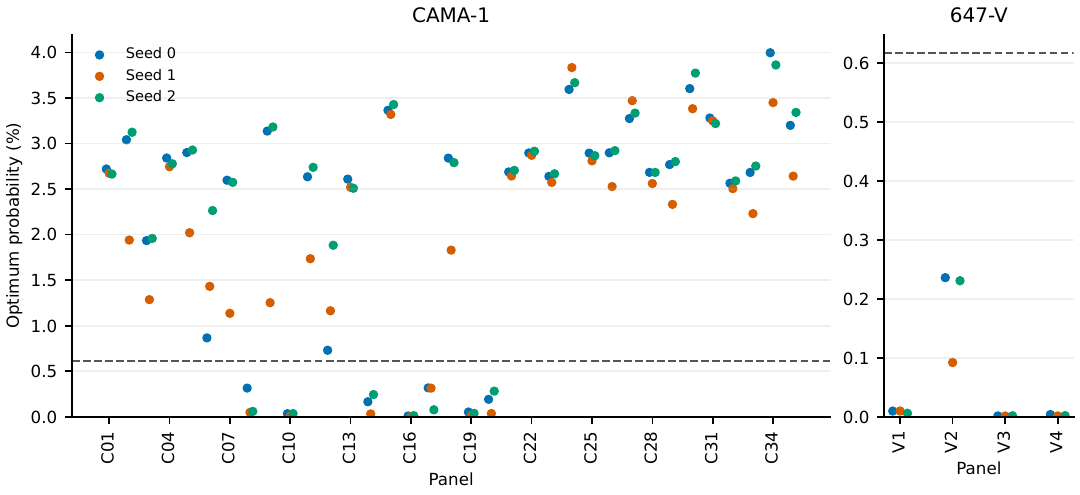}
\caption{Ideal penalty-QAOA optimum probability for all expanded panels and all three training seeds. The dashed line is uniform feasible sampling, at $100/162$ percent. Panel labels match Table~S11. Left: all CAMA-1 panels. Right: the four 647-V panels selected by measurement coverage.}
\end{figure}

\clearpage
\section{Reproducing the added experiments}
The added code and results are included in the reproducibility package. Run \texttt{prepare\_revision.py} to reproduce the frozen panel design from the supplied source tables. Run \texttt{revision\_experiments.py} with \texttt{--mode classical}, \texttt{--mode panels}, \texttt{--mode grover} and \texttt{--mode rescore} for the four computations. The first three modes produce independent result files. Rescoring uses the saved large-circuit feasible probabilities when available. Run \texttt{python -m unittest test\_revision -v} for numerical and search checks, then \texttt{build\_revision\_outputs.py} to regenerate the supplemental tables and panel figure. Script SHA-256 hashes are listed below; source-table hashes and panel membership are preserved in \texttt{evidence/revision\_design.json}.

\begin{itemize}

\item \texttt{prepare\_revision.py}\newline {\footnotesize\texttt{ad71abd57e05fd902c7a28cf3426c5a1\allowbreak e888fca6d9d23454ef1622b6c0b494e0}}.

\item \texttt{revision\_experiments.py}\newline {\footnotesize\texttt{232a3e45f773b6e3817dcd4a4f88f2f5\allowbreak 9f0940d6595bbc99bc080df60ab0e6f3}}.

\item \texttt{revision\_noiseless.py}\newline {\footnotesize\texttt{a5357753b6e8bbd86e2d941827f7cb2e\allowbreak 605eb321d1e67dd48a85765d1b3f41a7}}.

\item \texttt{test\_revision.py}\newline {\footnotesize\texttt{5acaa95aab14a7f15a9c538dc9ea5235\allowbreak e0b920dd7f33db5e8a69ee381b9eaf0a}}.

\item \texttt{test\_completed\_revision.py}\newline {\footnotesize\texttt{193ea836bd9373b2f38cb808f554f097\allowbreak 4c0286db043e6a7397a916d7438fa5e6}}.

\item \texttt{test\_noiseless\_revision.py}\newline {\footnotesize\texttt{8ddcfee1996c5f492483b8ae645b0ed1\allowbreak cf40e37c9e94c9b48eb3eba19bf1c73f}}.

\item \texttt{build\_revision\_outputs.py}\newline {\footnotesize\texttt{8d0f6d9e196eed0d86d38f312f6c9e11\allowbreak 989ef4d0298dcea4c1077411a029331b}}.

\item \texttt{rewrite\_prose.py}\newline {\footnotesize\texttt{ed6bc7c885976bd151baecb84dd7782e\allowbreak ef3092632a6c34e761aa403643cc7171}}.

\item \texttt{integrate\_revision.py}\newline {\footnotesize\texttt{4b76905a18f377473b76485c935ee015\allowbreak 7f0ba5c21b1203327fa892ecda702be7}}.

\item \texttt{integrate\_noiseless.py}\newline {\footnotesize\texttt{4892851fb7db5534a0a77f223e22614c\allowbreak 5df1838e2fcc793e8caa188394b39c21}}.

\item \texttt{revision.sbatch}\newline {\footnotesize\texttt{cbf13814e1782192d8c690891f6f4f2f\allowbreak d61b6a5ea91576e61b9fa70a26322cd2}}.

\item \texttt{noiseless\_revision.sbatch}\newline {\footnotesize\texttt{ba6ead793b8710fdc832b543ed4e8134\allowbreak 2aad3c61c369eacf3974a1553e6e1b2f}}.

\end{itemize}

\clearpage
\section{Complete task inventory}
Each item identifies one unique completed task by width, nonzero dose choices, UTC date/time in 2026 and task UUID. All 25 tasks are included in the analysis; the six original reference tasks are also listed here.
\begin{itemize}
\item 6q, $D=2$, 07-03 03:33: \texttt{24d41998-5edf-44bf-93a4-60fca948ae8c}.
\item 6q, $D=2$, 07-03 05:16: \texttt{98cb6dee-1d5d-4aec-9122-6a1c390a2608}.
\item 6q, $D=2$, 07-11 07:10: \texttt{c8ea4f7a-a64e-49d2-a7f8-f1ac1eff2136}.
\item 8q, $D=2$, 07-03 05:25: \texttt{4e307518-f350-4843-98ab-06f91315c190}.
\item 8q, $D=2$, 07-11 07:12: \texttt{e1a915b0-ff23-4769-8289-bfb7fcb0f74b}.
\item 9q, $D=3$, 07-03 05:33: \texttt{e88a4bac-5f06-4b2c-bcc8-69a20d23e527}.
\item 10q, $D=2$, 07-03 05:41: \texttt{9fcd57c3-97a5-4070-929d-d4ba5b7284f9}.
\item 10q, $D=2$, 07-11 07:19: \texttt{fad4ee6f-3815-4d92-af93-dbd2c19ec6a4}.
\item 12q, $D=2$, 07-11 07:25: \texttt{48d52888-7f37-4931-a135-8f831f1a208e}.
\item 12q, $D=3$, 07-03 05:51: \texttt{5cecf3e8-853e-448d-b296-d8bd2ef9f5e0}.
\item 14q, $D=2$, 07-11 07:38: \texttt{1ebf3086-bb62-48fc-9bdb-0df1dedd62cf}.
\item 15q, $D=3$, 07-03 06:03: \texttt{4d659093-99b5-4d58-9063-cd20724f7a98}.
\item 15q, $D=3$, 07-11 07:48: \texttt{566d8e0f-b67b-4786-86fa-3b112aaf15bd}.
\item 18q, $D=3$, 07-11 06:06: \texttt{4b60729f-56e1-45d5-a2d6-fec25d56c84c}.
\item 18q, $D=3$, 07-11 07:55: \texttt{b3fd2bd7-1b51-48e1-9c63-cbe7d55e1120}.
\item 18q, $D=3$, 07-12 13:37: \texttt{014e4aba-2301-4b48-87ba-28c46e97e84f}.
\item 20q, $D=5$, 07-11 08:09: \texttt{abdccbd0-9b84-4ba1-b5a7-91c6e0403be6}.
\item 20q, $D=5$, 07-12 13:47: \texttt{b4079656-2649-4163-9025-133821ff1bdd}.
\item 21q, $D=3$, 07-11 06:09: \texttt{fa0c9bef-c035-4232-a77b-75970696b227}.
\item 21q, $D=3$, 07-11 08:02: \texttt{30a44a65-2124-4664-9720-7d3d53e7498e}.
\item 21q, $D=3$, 07-12 13:40: \texttt{44849fe4-b8c2-4128-8792-ae5ef06009eb}.
\item 25q, $D=5$, 07-11 08:17: \texttt{e69c31e1-b34e-4869-bae9-217de933cb7d}.
\item 25q, $D=5$, 07-12 13:54: \texttt{b1386480-d348-4a01-ae1a-dbecf6d2a934}.
\item 30q, $D=5$, 07-11 08:23: \texttt{bb016dac-cabb-4e80-b568-0c0818917ab6}.
\item 35q, $D=5$, 07-11 08:33: \texttt{cfdf6f35-0141-46b3-a059-7f9e9fc85b9c}.
\end{itemize}

\clearpage
\section{Reproduction}
The analysis runs offline from the bundled records. From the analysis package directory, run \texttt{python reanalysis.py} to reproduce results from bundled evidence; then run \texttt{python build\_figures.py}. Run \texttt{python -m unittest discover -p test\_reanalysis.py -v} for circuit and scorer checks. Python, NumPy, matplotlib, and PennyLane versions are recorded in the package.

Run \texttt{python quality\_analysis.py} for the score distributions and greedy search, and \texttt{python sensitivity.py} for the frozen panel experiment. The latter preserves its frozen design and reports all seeds; reruns on a different platform can differ slightly because the optimization uses floating-point arithmetic. The original Pawsey output and its execution metadata are archived. Run \texttt{python build\_improvement\_outputs.py} for the added tables and figure, and \texttt{python -m unittest discover -p test\_improvements.py -v} for the additional independent checks.

Run \texttt{python analyze\_recovered\_hardware.py} to verify all downloaded tasks and reconstruct their observations. Then run \texttt{python build\_recovered\_outputs.py} to regenerate expanded tables and the repeat-execution plot. Run \texttt{python -m unittest discover -p test\_recovered\_hardware.py -v} to check enumeration through 35 qubits, reconciliation with the original six records, duplicate handling and circuit-based repetition grouping.

The separate directory \texttt{analysis/larger\_noiseless/} preserves the submitted package, completed execution archive, four result JSON files, Slurm logs and accounting record. The final batch script and code are in the completed archive. To recalculate a width, run \texttt{python larger\_noiseless.py --qubits 18 --output result.json}, substituting 20, 21 or 25 as appropriate. Run \texttt{python build\_larger\_outputs.py} to validate the saved outputs and generate Tables~S6--S7, then \texttt{python build\_figures.py} to regenerate the two-panel feasibility figure. The complete scientific test suite is \texttt{python -m unittest discover -p test\_*.py -v}.